\documentclass[
preprint,nofootinbib,amsmath,amssymb,aps,preprintnumbers,
]{revtex4-1}

\usepackage{graphicx}
\usepackage{bm}
\usepackage{braket}
\usepackage{cases} 
\usepackage{here}
\usepackage{color}
\usepackage{upgreek}
\usepackage{ulem}

\begin{document}

\preprint{KOBE-COSMO-20-07}

\title{A formalism for magnon gravitational wave detectors}

\author{Asuka Ito}
\email[]{asuka-ito@godzilla.kobe-u.ac.jp}

\author{Jiro Soda }
\email[]{jiro@phys.sci.kobe-u.ac.jp}

\affiliation{Department of Physics, Kobe University, Kobe 657-8501, Japan}


\begin{abstract}
In order to detect high frequency gravitational waves, we need a new detection method. 
In this paper, we develop a formalism for  a gravitational wave detector using magnons in a cavity. 
Using Fermi normal coordinates and taking the non-relativistic limit,
we obtain a Hamiltonian for magnons in gravitational wave backgrounds. 
Given the Hamiltonian, we show how to use the magnons for detecting high frequency gravitational waves.
Furthermore, as a demonstration of the magnon gravitational wave detector,
we give upper limits on GHz gravitational waves 
by utilizing known results of magnon experiments for an axion dark matter search.
\end{abstract}

\maketitle

\tableofcontents
\section{Introduction}
The discovery of gravitational waves by the interferometer detector LIGO in 2015 [1] opened up multi-messenger astronomy, where electromagnetic waves, gravitational waves, neutrinos, and cosmic rays are utilized to explore the universe. In the future, as the history of electromagnetic wave astronomy tells us, multi-frequency gravitational wave observations will be required to boost the multi-messenger astronomy.

It is useful to review the current status of gravitational wave observations~\cite{Kuroda:2015owv}.
Note that 
the lowest frequency we can measure is around $10^{-18}$ Hz,
below which the wavelength of gravitational waves 
exceeds the current Hubble horizon.
Measuring the temperature anisotropy and the B-mode polarization of the cosmic microwave background~\cite{Akrami:2018odb,Ade:2015tva}, we can probe gravitational waves with frequencies between $10^{-18}$ Hz and $10^{-16}$ Hz.
Astrometry of extragalactic radio sources is sensitive to gravitational waves with frequencies between $10^{-16}$ Hz and $10^{-9}$ Hz~\cite{Gwinn:1996gv,Darling:2018hmc}. 
The pulsar timing arrays, 
like EPTA~\cite{Lentati:2015qwp,Babak:2015lua},
IPTA~\cite{Perera:2019sca} and
NANOGrav~\cite{Arzoumanian:2018saf},
observe gravitational waves in the frequency band 
from $10^{-9}$ Hz to $10^{-7}$ Hz.
Doppler tracking of a space craft, which uses
a measurement method similar to the pulsar timing arrays, can search for gravitational waves 
in the frequency band from $10^{-7}$ Hz to $10^{-3}$ Hz~\cite{Armstrong:2003ay}.
The space interferometers LISA~\cite{AmaroSeoane:2012km} and DECIGO~\cite{Seto:2001qf} can cover the range
between $10^{-3}$ Hz and $10$ Hz.
The interferometer detectors LIGO~\cite{LIGO}, Virgo~\cite{Virgo}, and KAGRA~\cite{Somiya:2011np}  
with km size arm lengths
can search for gravitational waves with frequencies from $10$ Hz to $1$ kHz. 
In this frequency band, resonant bar experiments~\cite{Maggiore:1999vm} are
complementary to the interferometers~\cite{Acernese:2007ek}.
Furthermore,  interferometers  can be used to measure 
gravitational waves with
the frequencies between $1$ kHz and $100$ MHz.
Recently, a limit on gravitational waves at MHz was reported~\cite{Chou:2016hbb} and
a $0.75$ m arm length interferometer
gave an upper limit on $100$ MHz gravitational waves~\cite{Akutsu:2008qv}. 
At $100$ MHz, there is a waveguide experiment using an interaction between 
gravitational waves and electromagnetic fields~\cite{0264-9381-23-22-007}.
The interaction of gravitational waves with electromagnetic fields
is useful to explore high frequency gravitational waves and has been studied 
extensively~\cite{Li:2008qr,Li:2009zzy}.
Indeed, the interaction is utilized to constrain very high frequency gravitational waves higher than
$10^{14}$ Hz~\cite{Ejlli:2019bqj}.
Although  gravitational waves in the GHz range are theoretically  interesting~\cite{Ito:2019wcb},
no detector for GHz gravitational waves has been constructed.

In order to explore the GHz range, it would be useful to consider condensed matter systems. 
In our previous work, we pointed out that magnons in a cavity can be utilized to detect 
GHz gravitational waves~\cite{Ito:2019wcb}. There, we  gave observational constraints 
on GHz gravitational waves for the first time.
In this paper, we present the method in detail.
To treat the general coordinate invariance appropriately, we need to use Fermi
normal coordinates, or more precisely detector coordinates.
Furthermore, we study non-relativistic fermions to reveal the interaction 
between magnons and gravitational waves.
As a result,  we obtain a formalism for non-relativistic fermions
in curved spacetime, including a gravitational wave background as a special case.
Finally, as a demonstration, we will give upper limits on 
the spectral density of continuous gravitational waves ($95$ \% C.L.): 
$7.5 \times 10^{-19}  \  [{\rm Hz}^{-1/2}]$ at 14 GHz and
$8.7 \times 10^{-18}  \  [{\rm Hz}^{-1/2}]$ at 8.2 GHz, respectively,
by utilizing results of magnon experiments conducted recently~\cite{Crescini:2018qrz,Flower:2018qgb}.

The organization of the paper is as follows.
In section II, we study the Dirac equation in Fermi normal coordinates.
In section III, we take the non-relativistic limit to obtain a Hamiltonian
of the fermions.  
In section IV, we explain how to use magnons for detecting high frequency gravitational waves. 
Furthermore, we give upper limits on continuous gravitational waves in the GHz range.
The final section is devoted to the conclusion.
In Appendices A and B, we review how to derive Fermi normal coordinates and 
proper detector coordinates, respectively.
In particular in Appendix B, the reason why 
one can neglect gravity of the earth and use the Fermi normal coordinates as the proper detector frames approximately
will be clarified.
We also give a simple mathematical formula
for calculations in Appendix C.
\section{Dirac field in Fermi normal coordinates} \label{dirainncurve}
\quad
In order to study the effects of gravity on a fermion, 
we consider the Dirac equation in curved spacetime described by a metric $g_{\mu\nu}$.
It is given by
\begin{equation}
  i \gamma^{\hat{\alpha}} e^{\mu}_{\hat{\alpha}} \left( \partial_{\mu} - \Gamma_{\mu} 
                - i e A_{\mu} \right) \psi = m \psi \ ,  \label{dira}
\end{equation}
where $\gamma^{\hat{\alpha}}$,  $e$,  $m$, $A_{\mu}$ are the gamma matrices, the elementary charge, the mass of the fermion, 
 and the vector potential of $U(1)$ gauge theory, respectively.  
 The tetrad  $e^{\mu}_{\hat{\alpha}}$ satisfies 
\begin{equation}
e^{\hat{\alpha}}_{\mu} e^{\hat{\beta}}_{\nu} \eta_{\hat{\alpha}\hat{\beta}} = g_{\mu\nu} \ . \label{tetrad}
\end{equation}
Note that  $\eta_{\hat{\alpha}\hat{\beta}}$ is the Minkowski metric of a local inertial frame and
hat is used for the frame.
The spin connection is defined by
\begin{equation}
  \Gamma_{\mu} = 
   - \frac{i}{2} e^{\hat{\alpha}}_{\nu} \sigma_{\hat{\alpha}\hat{\beta}} 
        \left( \partial_{\mu} e^{\nu\hat{\beta}} + \Gamma^{\nu}_{\lambda\mu} e^{\lambda\hat{\beta}} \right), 
        \label{spicone}
\end{equation}
where
$\sigma_{\hat{\alpha}\hat{\beta}} = \frac{i}{4} [ \gamma_{\hat{\alpha}}, \gamma_{\hat{\beta}} ] $ is 
a generator of the Lorentz group and 
$\Gamma^{\mu}_{\nu\lambda}$ is the Christoffel symbol.

Since there is  the equivalence principle for gravity, 
the choice of  coordinates is quite important.
We should consider a proper reference frame, which coincides with 
the coordinates used in an experiment.
Actually, the proper reference frame can be approximated by Fermi normal coordinates 
(see the Appendix \ref{Fermi})
because the effects of the earth are negligible for our purposes as discussed in the Appendix \ref{EarthGra}.

In the Appendix \ref{Fermi},
we have derived an explicit expression of the metric in the Fermi normal coordinates as 
\begin{eqnarray}
  g_{00} &=& - 1 - R_{0i0j} x^{i} x^{j} \ ,  \label{met1} \\
  g_{0i} &=& -\frac{2}{3} R_{0jik} x^{j} x^{k} \ ,  \label{met2} \\
  g_{ij} &=& \delta_{ij} - \frac{1}{3} R_{ikjl} x^{k} x^{l} \ ,  \label{met3}
\end{eqnarray}
where the Riemann tensor is evaluated at $\bm{x} = 0$ and thus
it only depends on time $x^{0}$. 
Moreover, the inverse of the metric is approximately given by
\begin{eqnarray}
  g^{00} &=& - 1 + R^{0i0j} x_{i} x_{j}  \ ,   \label{met4}  \\
  g^{0i} &=& +\frac{2}{3} R^{0jik} x_{j} x_{k} \ ,   \label{met5}  \\
  g^{ij} &=& \delta_{ij} + \frac{1}{3} R^{ikjl} x_{k} x_{l} \ ,   \label{met6}
\end{eqnarray}
where we neglected higher order terms with respect to the curvature. 
From the metric (\ref{met1})-(\ref{met6}), one can obtain the Christoffel symbols:
\begin{eqnarray}
  \Gamma^{0}_{00} &=& 0 \ , \quad
  \Gamma^{0}_{0i} = R_{0i0j} x^{j} \ , \quad 
  \Gamma^{0}_{ij} = \frac{1}{3} \left( R_{0ijk} + R_{0jik} \right) x^{k} \ , \nonumber \\
  \Gamma^{i}_{00} &=& R_{0i0j} x^{j} \ , \quad
  \Gamma^{i}_{0j} = R_{0kji} x^{k} \ , \quad
  \Gamma^{i}_{jk} = \frac{1}{3} \left( R_{kijl} + R_{jikl} \right) x^{l} \ .  
 \label{chr}
\end{eqnarray}
%
%
The tetrad is constructed using the relation (\ref{tetrad}):
\begin{eqnarray}
  e^{\hat{\alpha}}_{0} &=& \delta^{\hat{\alpha}}_{0} 
                         - \frac{1}{2} \delta^{\hat{\alpha}}_{\alpha}  R^{\alpha}_{\ k0l}x^{k} x^{l} \ , \label{tet1} \\
  e^{\hat{\alpha}}_{i} &=& \delta^{\hat{\alpha}}_{i} 
                         - \frac{1}{6} \delta^{\hat{\alpha}}_{\alpha}  R^{\alpha}_{\ kil}x^{k} x^{l}c \ . \label{tet2} \\
  e^{0}_{\hat{\alpha}} &=& \delta^{0}_{\hat{\alpha}} +
                           \frac{1}{2} \delta^{0}_{\hat{\alpha}} R^{0}_{\ k0l}  -
                           \frac{1}{6} \delta^{j}_{\hat{\alpha}} R_{jk0l}   x^{k} x^{l}  \ ,  \label{tet3} \\              
  e^{i}_{\hat{\alpha}} &=& \delta^{i}_{\hat{\alpha}} -
                           \frac{1}{2} \delta^{ 0}_{\hat{\alpha}} R^{0\ i}_{\ k\ l} x^{k} x^{l} +
                           \frac{1}{6} \delta^{j}_{\hat{\alpha}} R^{i}_{\ kjl} x^{k} x^{l}     \ .   \label{tet4}
\end{eqnarray}
Substituting Eqs.\,(\ref{chr})-(\ref{tet4}) into Eq.\,(\ref{spicone}), 
we can evaluate the spin connection as
\begin{eqnarray}
  \Gamma_{0} &=&  -
  \frac{1}{2} \gamma^{\hat{0}} \gamma^{\hat{i}} R_{0i0j} x^{j}
                -
                 \frac{1}{4} \gamma^{\hat{i}} \gamma^{\hat{j}} R_{ij0k} x^{k} \ , \label{sc1} \\
  \Gamma_{i} &=& -
    \frac{1}{4} \gamma^{\hat{0}} \gamma^{\hat{j}} R_{0jik} x^{k} 
                -
                 \frac{1}{8} \gamma^{\hat{j}} \gamma^{\hat{k}} R_{jkil} x^{l}   \ .  \label{sc2}
\end{eqnarray}
Here we have rewritten $\delta_{\hat{\alpha}}^{\mu} \gamma^{\hat{\alpha}}$ as $\gamma^{\hat{\mu}}$ and 
we will do so throughout.

On the other hand,
the Dirac equation (\ref{dira}) can be rewritten as
\begin{eqnarray}
 i \gamma^{0} \partial_{0} \psi &=& \left[ i \gamma^{0} \left( \Gamma_{0} + i e A_{0} \right)
                                  - i \gamma^{j} \left( \partial_{j} - \Gamma_{j} - i e A_{j} \right)  
                                  +  m   \right]  \psi   \nonumber \\
                 &=&   \gamma^{0} H \psi \ ,
\end{eqnarray}
where we defined a Hamiltonian $H$ and the gamma matrices in curved spacetime
$\gamma^{\mu} = e^{\mu}_{\hat{\alpha}} \gamma^{\hat{\alpha}}$  
satisfying the relation  
\begin{equation}
  \{ \gamma^{\mu} ,  \gamma^{\nu} \}  =  - 2 g^{\mu\nu} \ .
\end{equation}
Let us express the Hamiltonian in terms of the gamma matrices of the local inertial frame
instead of those of curved spacetime.
Because of $\gamma^{0}\gamma^{0} = -g^{00}$, we obtain
\begin{equation}
  H =  (g^{00})^{-1}  \left[ i g^{00} \left( \Gamma_{0} + i e A_{0} \right)
                                  + i \gamma^{0}\gamma^{j} \left( \partial_{j} - \Gamma_{j} - i e A_{j} \right)  
                                  -  \gamma^{0} m   \right]  \ .   \label{hamihami}
\end{equation}
Using Eqs.\,(\ref{tet3}) and (\ref{tet4}), we calculate
\begin{eqnarray}
  \gamma^{0} \gamma^{j} &=&
        \left( e^{0}_{\hat{\alpha}} \gamma^{\hat{\alpha}}  \right)
        \left( e^{j}_{\hat{\beta}} \gamma^{\hat{\beta}}  \right)   \nonumber \\
        &=&  \left(  e^{0}_{\hat{0}} \gamma^{\hat{0}} + e^{0}_{\hat{a}} \gamma^{\hat{a}}  \right)
             \left(  e^{j}_{\hat{0}} \gamma^{\hat{0}}  + e^{j}_{\hat{b}} \gamma^{\hat{b}}  \right) \nonumber \\
        &=&  \gamma^{\hat{0}} \gamma^{\hat{j}} 
            - \frac{1}{2} \gamma^{\hat{0}} \gamma^{\hat{0}} R^{0\ j}_{\ k\ l} x^{k} x^{l} 
            + \frac{1}{6} \gamma^{\hat{0}} \gamma^{\hat{b}} R^{j}_{\ kbl}  x^{k} x^{l}  \nonumber \\
         && + \frac{1}{2} \gamma^{\hat{0}} \gamma^{\hat{j}}
              R^{0}_{\ k0l} x^{k}x^{l} 
            - \frac{1}{6} \gamma^{\hat{a}} \gamma^{\hat{j}}
              R_{ak0l} x^{k} x^{l}  \ .  \label{ga0j}
\end{eqnarray}
Together with Eq.\,(\ref{met4}), we have
\begin{eqnarray}
  (g^{00})^{-1}  \gamma^{0} \gamma^{j}  
       &\simeq&
            -  \gamma^{\hat{0}} \gamma^{\hat{j}} 
            - \frac{1}{2} R_{0kjl} x^{k} x^{l} 
            - \frac{1}{6} \gamma^{\hat{0}} \gamma^{\hat{a}} R_{jkal}  x^{k} x^{l}  \nonumber \\
         && - \frac{1}{2} \gamma^{\hat{0}} \gamma^{\hat{j}} R_{0k0l} x^{k}x^{l} 
            + \frac{1}{6} \gamma^{\hat{a}} \gamma^{\hat{j}} R_{ak0l} x^{k} x^{l}  \ .  \label{000j}
\end{eqnarray}
Similarly, one can obtain
\begin{equation}
  (g^{00})^{-1} \gamma^{0} \simeq -\gamma^{\hat{0}} - \frac{1}{2} \gamma^{\hat{0}} R_{0k0l} x^{k}x^{l}
                               + \frac{1}{6} \gamma^{\hat{a}} R_{ak0l} x^{k} x^{l} \ .   \label{000}
\end{equation}
Therefore, from Eqs.\,(\ref{hamihami}), (\ref{000j}) and (\ref{000}), the Hamiltonian 
expressed in the local inertial coordinates becomes
\begin{eqnarray}
 H &=&  i \Gamma_{0} + i \gamma^{\hat{0}} \gamma^{\hat{j}} \Gamma_{j} - eA_{0}   \nonumber \\
    &&   + 
    \bigg[  \gamma^{\hat{0}} \gamma^{\hat{j}}
         + \frac{1}{2} R_{0kjl} x^{k} x^{l}
         + \frac{1}{6} \gamma^{\hat{0}} \gamma^{\hat{a}} R_{jkal}  x^{k} x^{l}   \nonumber \\
    &&  \quad \quad + \frac{1}{2} \gamma^{\hat{0}} \gamma^{\hat{j}} R_{0k0l} x^{k}x^{l} 
         - \frac{1}{6} \gamma^{\hat{a}} \gamma^{\hat{j}} R_{ak0l} x^{k} x^{l}   \bigg]  
         \left( -i \partial_{j} -  eA_{j} \right) \nonumber \\
    &&  + \left[ 
         \gamma^{\hat{0}} + \frac{1}{2} \gamma^{\hat{0}} R_{0k0l} x^{k} x^{l} 
         - \frac{1}{6} \gamma^{\hat{a}} R_{ak0l} x^{k} x^{l}  
    \right]  m   \ .  \label{HHH}
\end{eqnarray}
Furthermore, substituting Eqs.\,(\ref{sc1}) and (\ref{sc2}) into the above Hamiltonian and 
rearranging terms, we have
\begin{eqnarray}
 H &=&  - 
 \frac{i}{2} \gamma^{\hat{0}} \gamma^{\hat{i}} R_{0i0j} x^{j}
        -
     \frac{i}{4} \gamma^{\hat{i}} \gamma^{\hat{j}} R_{0ikj} x^{k}
    -
      \frac{i}{8}  \gamma^{\hat{0}} \gamma^{\hat{i}} \gamma^{\hat{j}} \gamma^{\hat{k}} R_{jkil} x^{l}
     - eA_{0}   \nonumber \\
    &&   + 
    \bigg[  \gamma^{\hat{0}} \gamma^{\hat{i}} 
           \left( \delta^{j}_{i} + \theta^{j}_{i} \right)
         + \frac{1}{2} R_{0kjl} x^{k} x^{l}
         - \frac{1}{6} \gamma^{\hat{i}} \gamma^{\hat{j}} R_{ik0l} x^{k} x^{l}   \bigg]  
         \left( -i \partial_{j} -  eA_{j} \right) \nonumber \\
    &&  +  \gamma^{\hat{0}} \left[ 
          1 + \frac{1}{2} R_{0k0l} x^{k} x^{l} 
         - \frac{1}{6}  \gamma^{\hat{0}} \gamma^{\hat{i}} R_{ik0l} x^{k} x^{l}  
    \right]  m   \ ,  \label{HHH}
\end{eqnarray}
where we defined
\begin{equation}
  \theta^{j}_{i} = 
                \frac{1}{2} \delta^{j}_{i} R_{0k0l} x^{k}x^{l}
              + \frac{1}{6}  R_{jkil}  x^{k} x^{l}  \ .
\end{equation}

The Hamiltonian we have obtained  is the 4$\times$4 matrix including both  the particle and the anti-particle.
What we will consider is the non-relativistic fermion.
To take the non-relativistic limit of  the Hamiltonian of the fermion, 
we have to separate the particle and the anti-particle while expanding the Hamiltonian
in powers of $1/m$.
We will explicitly see how to perform this in the next section.
\section{Non-relativistic limit of Dirac equation}\label{nonrela}
In the previous section, we derived the Hamiltonian of 
a Dirac field in general curved spacetime with the Fermi normal coordinates.
Assuming that a fermion has a velocity well below the speed of light, which 
is the situation we will consider in the section \ref{GMreso},
we take the non-relativistic limit of the Hamiltonian.
The procedure in flat spacetime is known as the Foldy-Wouthuysen 
transformation~\cite{Foldy:1949wa,Bjorken:1965zz}.
We generalize it to the case of curved spacetime.

We first separate the Hamiltonian (\ref{HHH}) into the even part, the odd part and 
the terms multiplied by $m$ as
\begin{eqnarray}
 H &=&  - \frac{i}{2} \alpha^{i} R_{0i0j} x^{j}
        +
         \frac{i}{8}  \alpha^{i} \alpha^{j} \alpha^{k}  R_{jkil} x^{l}
         + \alpha^{i} \left(  \delta^{j}_{i}  +  \theta^{j}_{i} \right) \Pi_{j}  \nonumber \\
    &&   - eA_{0}  
         + \frac{i}{4} \alpha^{i} \alpha^{j} R_{0ikj} x^{k}
         + 
    \bigg[  
          \frac{1}{2} R_{0kjl} x^{k} x^{l}
         + \frac{1}{6} \alpha^{i} \alpha^{j}  R_{ik0l} x^{k} x^{l}   \bigg]  
         \Pi_{j} \nonumber \\
    &&  +  \left[ 
          \beta \left( 1 + \frac{1}{2} R_{0k0l} x^{k} x^{l} \right)
         - \frac{1}{6}  \beta \alpha^{i}  R_{ik0l} x^{k} x^{l}  \right]  m    \nonumber  \\
    &=&  \mathcal{O} + \mathcal{E} +  
          \left[ \beta \left( 1 + \frac{1}{2} R_{0k0l} x^{k} x^{l} \right)
                 - \frac{1}{6}  \beta \alpha^{i}  R_{ik0l} x^{k} x^{l}  \right]  m   \ ,
     \label{HHHH}
\end{eqnarray}
where we have defined $\beta = \gamma^{\hat{0}}$, $\alpha^{i} = \gamma^{\hat{0}} \gamma^{\hat{i}}$ and
$\Pi_{j} = -i \partial_{j} -  eA_{j}$ for brevity.
The even part, $\mathcal{E}$, means that the matrix has only block diagonal elements and 
the odd part, $\mathcal{O}$, means that the matrix has only block off-diagonal elements.
Any product of two even (odd) matrices is even and 
a product of even (odd) and odd (even) matrices becomes odd.
To take the non-relativistic limit of the Hamiltonian, 
we have to diagonalize the Hamiltonian (\ref{HHHH}) and expand the upper block diagonal part in powers of $1/m$.
More precisely, $1/m$ expansion is recognized as an expansion with respect to two parameters, $(m x)^{-1}$ and $v/c$.
Here, $x$ represents a typical length scale of 
the system which can be specified by the Fermi normal coordinates, 
i.e., $x \sim \sqrt{x^{i} x_{i}}$,
$v$ is the velocity of the fermion and $c$ denotes the speed of light.
Assuming $1/m x \ll 1 $ and $v/c \ll 1$, which hold in the situation of the section \ref{GMreso},
we will perform the $1/m$ expansion.
It is known that this can be done in flat spacetime
by repeating unitary transformations order by order in powers of $1/m$~\cite{Foldy:1949wa,Bjorken:1965zz}.
Let us generalize the method to the case of curved spacetime in the  Fermi normal coordinates.

We consider a unitary transformation,
\begin{equation}
  \psi' = e^{iS} \psi \ ,
\end{equation}
where $S$ is a time-dependent Hermitian 4 $\times$ 4 matrix.
Observing that
%
%
\begin{eqnarray}
  i \frac{\partial \psi'}{\partial t} &=& i \frac{\partial}{\partial t} \left( e^{iS} \psi \right) \nonumber \\
               &=& e^{iS} \left( i \frac{\partial \psi}{\partial t} \right) 
                   + i \left( \frac{\partial}{\partial t} e^{iS} \right) \psi   \nonumber \\
               &=& \left[ e^{iS} H e^{-iS} +  i \left( \frac{\partial}{\partial t} e^{iS} \right)  e^{-iS} \right]
                     \psi'  \ ,
\end{eqnarray}
we find that the Hamiltonian after the unitary transformation is given by
\begin{equation}
  H' = e^{iS} H e^{-iS} +  i \left( \frac{\partial}{\partial t} e^{iS} \right)  e^{-iS} \ .  \label{traH}
\end{equation}
We now assume that $S$ is proportional to powers of $1/m$ and
expand the transformed Hamiltonian (\ref{traH}) 
in powers of $S$ up to the order of $1/m$.
Using Eqs.\,(\ref{cam}) and (\ref{expanS}) in Eq.\,(\ref{traH}),
we obtain
\begin{eqnarray}
  H' &=& H + i \big[ S,H \big] - \frac{1}{2} \big[ S,\big[S,H \big] \big] 
         - \frac{i}{6} \big[S,\big[S, \big[S , H \big] \big] \big]  + \cdots \nonumber \\
     &&  - \dot{S} - \frac{i}{2} \big[ S,\dot{S} \big] 
         + \cdots \ .  \label{newH}
\end{eqnarray}

First, let us eliminate the off-diagonal part of the Hamiltonian (\ref{HHHH}) at the order of $m$ by a
unitary transformation.
Then we will drop the higher order terms with respect to 
the Riemann tensor, which only depends on time, and 
derivatives of the Riemann tensor with respect to the time by assuming that 
they are small enough%
\footnote{Then, the Hermiticity of the non-relativistic Hamiltonian 
is guaranteed~\cite{Huang:2008kh}.
}.

To cancel the last term in the square bracket of (\ref{HHHH}),
we take
\begin{equation}
  S = - \frac{i}{2m}\beta \left( - \frac{1}{6} \beta \alpha^{i} R_{ik0l} x^{k} x^{l} m \right) \ .
\end{equation}
We then obtain
\begin{eqnarray}
  i \big[ S,H \big] &\simeq& \frac{1}{6} \beta \alpha^{i} R_{ik0l} x^{k}x^{l} m
                      - \frac{1}{12}  \big[ \alpha^{i} , \alpha^{j} \big]
                         R_{ik0l} x^{k} x^{l}  \Pi_{j}  \nonumber \\
                    &&+ \frac{i}{6} \alpha^{i}\alpha^{j} R_{0ikj} x^{k} 
                      + \frac{i}{12} \alpha^{i}\alpha^{j} R_{0jik} x^{k}  \ . \label{SH}
\end{eqnarray}
Therefore, from Eqs.\,(\ref{newH}) and (\ref{SH}), we have the transformed Hamiltonian as
%
%
%
%
%
%
%
%
%
%
\begin{eqnarray}
 H' &\simeq& H + i \big[ S,H \big] \nonumber \\
    &\simeq&  -  \frac{i}{2} \alpha^{i} R_{0i0j} x^{j}
         +
          \frac{i}{8}  \alpha^{i} \alpha^{j} \alpha^{k}  R_{jkil} x^{l}
         + \alpha^{i} \left(  \delta^{j}_{i}  +  \theta^{j}_{i} \right) \Pi_{j}  \nonumber \\
    &&    - eA_{0}
          + \frac{i}{6}   R_{0iki} x^{k}  
          + 
          \frac{2}{3} R_{0kil} x^{k} x^{l} \Pi_{i} 
     -\frac{i}{8} [\alpha^{i},\alpha^{j}] 
                        R_{ijk0} x^{k}
          \nonumber \\
    &&  +  
          \beta \left( 1 + \frac{1}{2} R_{0k0l} x^{k} x^{l} \right)
           m    \nonumber  \\
    &=&  \mathcal{O} + \mathcal{E}' +  
           \beta \left( 1 + \frac{1}{2} R_{0k0l} x^{k} x^{l} \right)
                   m   \ ,
     \label{HHHHH}
\end{eqnarray}
where we have used the relation $\big\{ \alpha^{i} ,\alpha^{j} \big\} = 2 \delta^{ij}$.
One can see that only even terms remain at the order of $m$, as expected.

Next, we focus on the order of $m^{0}$ and eliminate the odd terms by a unitary transformation.
In order to do so, we choose the Hermitian operator to be
\begin{equation}
  S' = -\frac{i}{2m} \beta \left( 
            \mathcal{O}  -\frac{1}{2} \alpha^{j} R_{0k0l} x^{k} x^{l} \Pi_{j}
                         + \frac{i}{2} \alpha^{j} R_{0k0j} x^{k}   \right)   \ .
\end{equation}
It is straightfoward to obtain
\begin{eqnarray}
  i \big[ S',H' \big] &\simeq&  - \mathcal{O} + \frac{1}{m} \beta \mathcal{O}^{2} 
                                + \frac{1}{2m} \beta \big[ \mathcal{O}, \mathcal{E}' \big] \nonumber \\
          &&   - \frac{1}{2m} \beta \alpha^{i} \alpha^{j} R_{0k0l} x^{k} x^{l} \Pi_{i} \Pi_{j}  
               + \frac{i}{m} \beta R_{0k0i} x^{k}  \Pi_{i}   \nonumber \\
          &&   + \frac{i}{4m} \beta \big[ \alpha^{i} ,\alpha^{j} \big] R_{0k0i} x^{k} \Pi_{j}  
               + \frac{1}{4m} \beta  R_{0i0i}    \nonumber \\
          &&   - \frac{i}{4m} \beta \alpha^{j} R_{0k0l} x^{k} x^{l} \left( \partial_{j} e A_{0} \right) \ , 
          \label{S,H}
\end{eqnarray}
Furthermore, up to the order of $1/m$, we can deduce
\begin{eqnarray}
  - \frac{1}{2} \big[ S',\big[S',H' \big] \big]  &\simeq&
      - \frac{1}{2} \big[ S',  i \mathcal{O} \big]  \nonumber \\
      &\simeq& 
      - \frac{1}{2m} \beta \mathcal{O}^{2} 
      + \frac{1}{4m} \beta \alpha^{i} \alpha^{j}  R_{0k0l} x^{k} x^{l} \Pi_{i} \Pi_{j}
      - \frac{i}{2m}  \beta  R_{0k0i} x^{k} \Pi_{i}
      \nonumber \\
   && - \frac{i}{8m} \beta \big[ \alpha^{i}, \alpha^{j} \big] R_{0k0i} x^{k} \Pi_{j}
      - \frac{1}{8m} \beta  R_{0i0i} \ , 
\end{eqnarray}
and
\begin{equation}
  - \dot{S}' \simeq 
      \frac{i}{2m} \beta  \dot{\mathcal{O}}  
    + \frac{i}{4m} \beta \alpha^{j} R_{0k0l} x^{k} x^{l} e \dot{A}_{j}  \ .
\end{equation}
Therefore, the Hamiltonian after the unitary transformation is given by
\begin{eqnarray}
  H''  &\simeq& H' + i \big[ S',H' \big] - \frac{1}{2} \big[ S',\big[S',H' \big] \big] - \dot{S}'  \nonumber \\
  &\simeq&  
    - \frac{i}{4m} \beta \alpha^{j} R_{0k0l} x^{k} x^{l} e E_{j}
    + \frac{1}{2m} \beta \left(  \big[ \mathcal{O}, \mathcal{E}' \big] + i \dot{\mathcal{O}} \right) \nonumber \\
  && + \mathcal{E}'
     + \frac{1}{2m} \beta \mathcal{O}^{2} 
     - \frac{1}{4m} \beta  \alpha^{i} \alpha^{j} R_{0k0l} x^{k} x^{l} \Pi_{i} \Pi_{j}  \nonumber \\
  && 
     + \frac{i}{2m} \beta R_{0k0i} x^{k}  \Pi_{i}
     + \frac{i}{8m} \beta \big[ \alpha^{i} ,\alpha^{j} \big] R_{0k0i} x^{k} \Pi_{j} 
     + \frac{1}{8m} \beta  R_{0i0i}   \nonumber \\
  && + \beta \left( 1 + \frac{1}{2} R_{0k0l} x^{k} x^{l} \right)  m   \nonumber \\
  &=&  \mathcal{O}' + \mathcal{E}'' + \beta \left( 1 + \frac{1}{2} R_{0k0l} x^{k} x^{l} \right)  m \ ,
\end{eqnarray}
where $E_{j} \equiv \partial_{j} A_{0} -  \dot{A}_{j}$ is an electric field.
We see that $\mathcal{O}'$ has only terms of order of $1/m$, so that 
odd terms at the order of $m^{0}$ have been eliminated.

Finally, we will eliminate the odd term $\mathcal{O}'$ and then 
the Hamiltonian will consist of only even terms up to the order of $1/m$, which we want to get.
To this end, 
we now choose the Hermitian operator of a unitary transformation as
\begin{equation}
  S'' = -\frac{i}{2m} \beta \left( 
            \mathcal{O}'  - \frac{i}{4m} \beta \alpha^{i} e E_{i} R_{0k0l} x^{k} x^{l}  \right)  \ .
\end{equation}
Then, up to the order of $1/m$, we have
\begin{equation}
   i \big[ S'' , H'' \big] \simeq  - \mathcal{O}'  \ .
\end{equation}
Therefore, we have the transformed Hamiltonian as
\begin{eqnarray}
  H''' &\simeq&  H''  +  i \big[ S'' , H'' \big]   \nonumber \\
       &\simeq&  \mathcal{E}'' + \beta \left( 1 + \frac{1}{2} R_{0k0l} x^{k} x^{l} \right)  m \ , \label{H'''}
\end{eqnarray}
where $\mathcal{E}''$ is given by
\begin{eqnarray}
   \mathcal{E}'' &=& 
     - eA_{0}   
     + \frac{i}{6}   R_{0iki} x^{k}  
     + \frac{2}{3} R_{0kil} x^{k} x^{l} \Pi_{i}
     -\frac{i}{8} [\alpha^{i},\alpha^{j}] R_{ijk0} x^{k}
     + \frac{1}{2m} \beta \mathcal{O}^{2} 
     - \frac{1}{4m} \beta  \alpha^{i} \alpha^{j} R_{0k0l} x^{k} x^{l} \Pi_{i} \Pi_{j} \nonumber \\
  && + \frac{i}{2m} \beta R_{0k0i} x^{k}  \Pi_{i}
     + \frac{i}{8m} \beta \big[ \alpha^{i} ,\alpha^{j} \big] R_{0k0i} x^{k} \Pi_{j} 
     + \frac{1}{8m} \beta  R_{0i0i}   \label{E''}  \ .
\end{eqnarray}
Moreover, the fourth term in the first line of Eq.\,(\ref{E''}) can be evaluated as
\begin{eqnarray}
  \frac{1}{2m} \beta \mathcal{O}^{2} &\simeq&
     \frac{i}{8m}\beta  \big[ \alpha^{i} ,\alpha^{j} \big] \epsilon_{ilm} e B^{m}
     \left(  \delta_{lj}  + 2  \theta_{lj}  \right)  \nonumber \\
   && -\frac{i}{4m}\beta \big[ \alpha^{i} ,\alpha^{j} \big]
     \left( \frac{1}{4} R_{lmji}  +  \delta^{l}_{j}R_{0i0m} \right) x^{m} \Pi_{l} \nonumber \\
   && + \frac{1}{2m}\beta \left( \delta_{ij} + 2 \theta_{ij} \right) \Pi_{i} \Pi_{j} 
      + \frac{i}{12 m}\beta R_{kikj} x^{j} \Pi_{i}  
      + \frac{1}{4m} \beta R_{0i0i}   \nonumber \\
   && + \frac{1}{16m} \beta \alpha^{i} \alpha^{j} \alpha^{k} \alpha^{l} R_{ijkl}
      - \frac{i}{16m} \beta \big\{ \alpha^{i} , \alpha^{j} \alpha^{k} \alpha^{l} \big\}
        R_{kljm} x^{m} \Pi_{i} 
        \ ,  \label{Ono2}
\end{eqnarray}
%
%
where $B^{i} \equiv \frac{1}{2} \epsilon^{ijk} (\partial_{j}A_{k} - \partial_{k}A_{j})$ is a magnetic field.
Using Eqs.\,(\ref{E''}), (\ref{Ono2}) and the relation, 
$\big[ \alpha^{i} ,\alpha^{j} \big] = 2 i \epsilon_{ijk} \sigma^{k}$,
in the transformed Hamiltonian (\ref{H'''}),
we finally arrive at the Hamiltonian for a non-relativistic fermion up to the order of $1/m$ as 
\begin{eqnarray}
  H''' &=& 
   \left( 1 + \frac{1}{2} R_{0k0l} x^{k} x^{l} \right)  m  
     - eA_{0}
     + \frac{i}{6}   R_{0iki} x^{k}  
     + \frac{2}{3} R_{0kil} x^{k} x^{l} \Pi_{i}
     + \frac{1}{4} \epsilon_{ijl} \sigma^{l} R_{ijk0} x^{k}
       \nonumber \\
 &&  + \frac{1}{2m} 
            \left[ \delta_{ij} \left( 1 +  \frac{1}{2} R_{0k0l} x^{k}x^{l} \right)
              + \frac{1}{3}  R_{jkil}  x^{k} x^{l} \right] \Pi_{i} \Pi_{j} \nonumber \\
 &&  - \frac{e}{2m} \sigma^{i} B^{j} \left[  \delta_{ij} 
       \left( 1 + \frac{1}{2} R_{0k0l} x^{k} x^{l} + \frac{1}{6} R_{mkml} x^{k} x^{l}  \right)
              - \frac{1}{6}  R_{ikjl} x^{k} x^{l}  \right] \nonumber \\
 &&  + \frac{1}{8m} \epsilon_{ijk} \sigma^{k} 
       \left(  R_{ijlm}  + 2 \delta_{jm} R_{0i0l} \right) x^{l} \Pi_{m}  \nonumber \\
 &&  + \frac{1}{8 m} \left( 3 R_{0i0i} -  R_{ijij} \right) 
     + \frac{i}{2 m} \left( R_{0i0j}  - \frac{1}{3} R_{kikj}  \right) x^{i} \Pi_{j}  \ . \label{Hddd}
\end{eqnarray}
We mention that 
the non-relativistic Hamiltonian up to the order of $m$ was derived 
in~\cite{Parker:1980kw} and coincides with our result.
The first term is the rest mass and its correction from the gravity at a point $x^{i}$.
The third term represents gravitational redshift, namely energy shift due to gravity.
The first term in the last line gives the same effect at the order of $1/m$.
We find that 
the fourth and the sixth terms describe gravitational effects on the motion of a particle where
they contains the time derivative of the curvature, which has been assumed to be small.
The second term in the last line also
gives the similar effect at the order of $1/m$.
The fifth term is a gravitational effect on a spin.
However, this does not affect magnons because an spatial integration
over a ferromagnetic sample becomes zero as we will see later.
The third line represents interactions between gravity and a 
spin in the presence of an external magnetic field.
This is what causes the spin resonance and/or the excitation of magnons as we will see in the next section.
The fourth line is a spin-orbit coupling mediated by gravity.

In vacuum, the Riemann tensor coincides with the Weyl tensor.
Then it may be useful to rewrite the Riemann tensor of the Hamiltonian (\ref{Hddd}) 
in terms of the electric $E_{ij}$
 and magnetic $H_{ij}$ components of the Weyl tensor $C_{\mu\nu}^{\ \ \rho\sigma}$, defined by
\begin{equation}
  R_{\mu\nu}^{\ \ \rho\sigma} = C_{\mu\nu}^{\ \ \rho\sigma} 
    = 4 \left( \gamma^{[\rho}_{[\mu} - \delta^{|0|}_{[\mu} \delta^{[\rho}_{|0|}  \right) E^{\sigma]}_{\nu]}
      + 2 \epsilon_{\mu\nu\alpha 0} \delta^{[\rho}_{0} H^{\sigma]\alpha} 
      + 2 \epsilon^{\rho\sigma\alpha 0} \delta^{0}_{[\mu} H_{\nu]\alpha} \ ,
\end{equation}
where $\gamma_{\mu\nu}$ is an induced three dimensional metric, i.e., 
$\gamma_{00} = \gamma_{0i} = 0, \gamma_{ij} \simeq \delta_{ij}$.
Substituting the above relation into the Hamiltonian (\ref{Hddd}),
we  obtain
\begin{eqnarray}
  H''' &=& 
   \left( 1 + \frac{1}{2} E_{kl} x^{k} x^{l} \right)  m  
     - eA_{0}
     + \frac{i}{6}   H_{il}\epsilon_{kil} x^{k}  
     + \frac{2}{3} H_{km} \epsilon_{ilm} x^{k} x^{l} \Pi_{i}  
     - \frac{1}{2} \sigma^{l} H_{kl} x^{k}
     \nonumber\\
 &&    + \frac{1}{2m} \left[ \delta_{ij} \left( 1 + \frac{1}{2} E_{kl} x^{k} x^{l}  \right)   
 + \frac{4}{3}   \delta_{[j|[i} E_{l]|k]} x^{k} x^{l}  \right] \Pi_{i}\Pi_{j}
  \nonumber \\
 &&  - \frac{e}{2m} \sigma^{i} B^{j} \left[  \delta_{ij} 
       \left( 1 + \frac{1}{2} E_{kl} x^{k} x^{l} + \frac{2}{3}  \delta_{[m|[m} E_{l]|k]}  x^{k} x^{l}  \right)
              - \frac{2}{3}   \delta_{[i|[j} E_{l]|k]} x^{k} x^{l}  \right] \nonumber \\
 &&  + \frac{1}{4m} \epsilon_{ijk} \sigma^{k} 
       \left(  2 \delta_{[i|[l} E_{m]|j]}  +  \delta_{jm} E_{il} \right) x^{l} \Pi_{m}  \nonumber \\
 &&  + \frac{1}{2m} \left( \frac{3}{4} E_{ii} -  \delta_{[i|[i} E_{j]|j]} \right) 
     + \frac{i}{2m} \left( E_{ij}  - \frac{4}{3} \delta_{[k|[k} E_{j]|i]}  \right) x^{i} \Pi_{j}  \ . \label{HWeyl}
\end{eqnarray}

Although the expression (\ref{Hddd}) is applicable to a general curved spacetime, 
let us focus on gravitational waves as gravitational effects from now on.
The Riemann tensor for a perturbed metric 
$g_{\mu\nu} = \eta_{\mu\nu} + h_{\mu\nu}$ at the linear order is given by
\begin{equation}
    R^{\alpha}{}_{\mu\beta\nu} 
             =  \frac{1}{2}(h^{\alpha}_{\ \nu,\mu\beta}-h_{\mu\nu\ \beta}^{\ \ ,\alpha}
                   -h^{\alpha}_{\ \beta,\mu\nu}+h_{\mu\beta\ ,\nu}^{\ \ ,\alpha}) \ , \label{Rie2}
\end{equation}
where $\eta_{\mu\nu}$ stands for a flat spacetime metric and $h_{\mu\nu}$ represents a deviation from the flat spacetime.
Because the Riemann tensor (\ref{Rie2}) is invariant under gauge transformations,
we can use any coordinate to evaluate the Riemann tensor included in Eq.\,(\ref{Hddd}).
We then take the transverse traceless gauge, i.e., $h_{0\mu} = h_{ii} = h_{ij,j} = 0$.
As a result, one can obtain 
\begin{eqnarray}
  R_{0i0j} &=& - \frac{1}{2} \ddot{h}_{ij}  \ , \nonumber \\
  R_{0ijk} &=& \frac{1}{2}  \left( \dot{h}_{ij,k} - \dot{h}_{ik,j} \right) \ , \nonumber \\
  R_{ijkl} &=& \frac{1}{2}  \left( h_{il,jk} + h_{jk,il} - h_{jl,ik} - h_{ik,jl} \right)   \label{RieGW} \ .
\end{eqnarray}
Note that they are evaluated at the origin, $x^{i} = 0$, so that 
they do not depend on spatial coordinates.
Substituting (\ref{RieGW}) into (\ref{Hddd}), we finally obtain
\begin{eqnarray}
  H''' &=& 
   \left( 1 - \frac{1}{4} \ddot{h}_{ij} x^{i} x^{j} \right)  m  
     - eA_{0}
   + \frac{1}{3} \left( h_{ki,l} - h_{kl,i} \right) x^{k} x^{l} \Pi_{i}  
   - \frac{1}{4} \epsilon_{ijl} \sigma^{l} \dot{h}_{ki,j} x^{k}
   \nonumber\\
 &&    + \frac{1}{2m} \left[ 
         \delta_{ij} \left( 1 - \frac{1}{4} \ddot{h}_{kl} x^{k} x^{l}  \right) 
             + \frac{1}{6} 
            \left(  h_{jl,ki} + h_{ki,jl}  - h_{kl,ij} - h_{ij,kl} \right) 
             x^k x^l \right] \Pi_{i} \Pi_{j}
  \nonumber \\
 &&  - \frac{e}{2m} \sigma^{i} B^{j} \left[  \delta_{ij} 
       \left( 1 - \frac{1}{3} \ddot{h}_{kl} x^{k} x^{l}   \right)
           - \frac{1}{12} 
           \left( h_{il,kj} + h_{kj,il} - h_{kl,ij} - h_{ij,kl} \right)  x^{k} x^{l}  \right] \nonumber \\
 &&  + \frac{1}{8m} \epsilon_{ijk} \sigma^{k} 
       \left(  h_{im,jl}  - h_{il,jm} 
         -  \delta_{jm} \ddot{h}_{il} \right) x^{l} \Pi_{m}  
     - \frac{i}{6m} \ddot{h}_{ij} x^{i} \Pi_{j}  \ ,  \label{Hddd2}
\end{eqnarray}
where we have used the equation of motion for gravitational waves, i.e., $\Box h_{ij} = 0$.

In the next section, 
we will see that gravitational waves excite magnons, which are collective excitation of spins
through the interaction in the third line in Eq.\,(\ref{Hddd2}).
\section{Magnon gravitational wave detectors} \label{GMreso}
In the section \ref{nonrela},
we revealed gravitational effects on a non-relativistic Dirac fermion
 in the   Fermi normal coordinates.
As you can see in Eq.\,(\ref{Hddd2}), 
if one consider a freely falling point particle and set  Fermi normal coordinates, 
the particle does not feel perturbative gravity $h_{ij}$ at the origin because of the equivalence principle.
However, gravitational effects are canceled, of course, only at one point and thus
an object with finite dimension feels gravitation.
In the case of magnons, we prepare, for example,
a ferromagnetic sample in an external magnetic field and then 
the sample feels gravity since it has finite size.
Thus, magnons can be excited by gravitational waves.
To examine the effect of gravitational waves on magnons,
it is appropriate to set a Fermi normal coordinate with the origin
 placed at the center 
of the ferromagnetic sample.
Then we can use the discussion of the section \ref{nonrela}.

We consider a ferromagnetic sample in an external magnetic field.
Such a system is described by the Heisenberg model~\cite{Heisenberg1926}:
\begin{eqnarray}
  H_{{\rm spin}} =   - 2\mu_{B} B_{z} \sum_{i}   \hat{S}^{z}_{(i)} 
       -  \sum_{i,j} J_{ij} \hat{\bm{S}}_{(i)} \cdot \hat{\bm{S}}_{(j)} \ ,  \label{heis3} 
\end{eqnarray}
where the Bohr magneton $\mu_{B} = e/2m_{e}$ is defined by 
the elementary charge $e$ and the mass of electrons $m_{e}$.
We applied an external magnetic field along the $z$-direction, $B_{z}$,
without loss of generality because of isotropy.
Here, $i$ specifies each site of spins.
The first term is the conventional Pauli term, which turns the spin direction to be along
the external magnetic field.
The second term represents the exchange interactions
between spins with the strength $J_{ij}$.

Next, we take into account the effect of gravitational waves on the system.
From Eq.\,(\ref{Hddd2}), the interaction Hamiltonian between gravitational waves and 
a spin in the ferromagnetic sample is
\begin{equation}
  H_{{\rm GW}} =
  - \mu_{B} B_{a}  \hat{S}^{b}_{(i)}  Q_{ab} \ ,  \label{qab}
\end{equation}  
where we have defined 
\begin{equation}
  Q_{ij} =   
        - \frac{2}{3} \delta_{ij}  \ddot{h}_{kl}|_{\bm{x}=0} \, x^{k} x^{l}   
           - \frac{1}{6} 
           \left( h_{il,kj} + h_{kj,il} - h_{kl,ij} - h_{ij,kl} \right)|_{\bm{x}=0} \,
             x^{k} x^{l} \ .
\end{equation}
Note that we neglected 
the fourth term in Eq\,(\ref{Hddd2}) because its 
integration over the spins becomes zero (see Eq.\,(\ref{integration})).
It represents the effect of gravitational waves 
on a spin located at $x^{i}$ in the Fermi normal coordinates. 
Indeed, at the origin, $x^{i}=0$, we see that $Q_{ij}=0$.
From Eqs.\,(\ref{heis3}) and (\ref{qab}), the total Hamiltonian of the system is
\begin{eqnarray}
  H_{{\rm tot}}  &=&  H_{{\rm spin}} + H_{{\rm GW}} \nonumber \\
       &=&
       - \mu_{B} \left( 2 \delta_{za} + Q_{za} \right) 
          B_{z} \sum_{i}   \hat{S}^{a}_{(i)} 
       -  \sum_{i,j} J_{ij} \hat{\bm{S}}_{(i)} \cdot \hat{\bm{S}}_{(j)} \ .  \label{total}
\end{eqnarray}

The spin system (\ref{total}) can be rewritten by using
the Holstein-Primakoff transformation~\cite{Holstein:1940zp}: 
\begin{eqnarray}
  \begin{cases}
   \hat{S}_{(i)}^{z} = \frac{1}{2} - \hat{C}_{i}^{\dagger} \hat{C}_{i} \ , & \\
   \hat{S}_{(i)}^{+} = \sqrt{1- \hat{C}_{i}^{\dagger} \hat{C}_{i}} \ \hat{C}_{i}  \ , & \\
   \hat{S}_{(i)}^{-} = \hat{C}_{i}^{\dagger} \sqrt{1- \hat{C}_{i}^{\dagger} \hat{C}_{i}} \ , &
  \end{cases} \label{prim}
\end{eqnarray}
where bosonic operators $\hat{C}_{i}$ and $\hat{C}^{\dagger}_{i}$ 
satisfy commutation relations $[\hat{C}_{i}, \hat{C}^{\dagger}_{j}]=\delta_{ij}$ and
$\hat{S}_{(j)}^{\pm} = \hat{S}_{(j)}^{x} \pm i \hat{S}_{(j)}^{y}$ are the ladder operators.
It is easy to check that the SU(2) algebra, $[\hat{S}^{i} , \hat{S}^{j}] = i \epsilon_{ijk} \hat{S}^{k}$ 
($i,j,k = x, y, z$), is satisfied even after the transformation (\ref{prim}).
We note that $\hat{C}_{i}^{\dagger} \hat{C}_{i}$ represents the particle numbers of the boson
created by the creation operator $\hat{C}_{i}^{\dagger}$.
The bosonic operators describe spin waves with dispersion relations determined by 
$B_{z}$ and $J_{ij}$. 
Furthermore, provided that contributions from the surface of the sample are negligible,
one can expand the bosonic operators by plane waves as 
\begin{equation}
  \hat{C}_{i} = \sum_{\bm{k}} \frac{e^{-i\bm{k} \cdot \bm{r}_{i}}}{\sqrt{N}} \hat{c}_{k} \ , \label{bbb}
\end{equation}
where $\bm{r}_{i}$ is the position vector of the $i$ spin.
The excitation of the spin waves created by $\hat{c}_{k}^{\dagger}$ is called a magnon.

We now rewrite the spin system (\ref{total})
by magnons with the Holstein-Primakoff transformation (\ref{prim}) and 
then we only focus on the homogeneous mode of magnons, i.e., $k = 0$ mode. 
Then, the second term in the total Hamiltonian (\ref{total}) is irrelevant because
it does not contribute to the homogeneous mode.
Furthermore, because $Q_{zz}$ does not contribute to the resonance of the spins, namely excitation of magnons,
we will drop it. 
Thus we have
\begin{equation}
  H_{{\rm tot}} =      
        \mu_{B}  B_{z} \sum_{i}  
      \left[  2 \hat{C}_{i}^{\dagger} \hat{C}_{i}  + 
      \frac{\hat{C}_{i} + \hat{C}_{i}^{\dagger}}{2}  Q_{zx}
    + \frac{\hat{C}_{i} - \hat{C}_{i}^{\dagger}}{2i} Q_{zy} \right]  \ . 
    \label{toHami}
\end{equation}

Now let us consider a planar gravitational wave propagating in the $z$-$x$ plane, namely,
the wave number vector of the gravitational wave $\bm{k}$ has a direction 
$\hat{k} = ( \sin\theta , 0 , \cos\theta )$.
Moreover, we postulate that the wavelength of the gravitational wave is much longer than the dimension 
of the sample and it is necessary for the validity of the Fermi normal coordinates.
This situation is actually satisfied in the case of usual cavity experiments for magnons.
We can expand the gravitational wave $h_{ij}$ in terms of linear polarization tensors satisfying 
$e^{(\sigma)}_{ij} e^{(\sigma')}_{ij} = \delta_{\sigma\sigma'}$ as
\begin{equation}
  h_{ij}(\bm{x}, t) = h^{(+)}(\bm{x}, t) e^{(+)}_{ij} 
           + h^{(\times)}(\bm{x}, t) e^{(\times)}_{ij} \ . \label{expan}
\end{equation}
%
More explicitly, we took the representation 
\begin{numcases}
  {}
  h^{(+)}(\bm{x}, t) = \frac{h^{(+)}}{2} 
      \left( e^{-i(w_{h}t - \bm{k}\cdot\bm{x})} 
           + e^{i(w_{h}t - \bm{k}\cdot\bm{x})} \right) \ , & \\
  h^{(\times)}(\bm{x}, t) = \frac{h^{(\times)}}{2} 
        \left( e^{-i(w_{h}t - \bm{k}\cdot\bm{x} + \alpha)} 
             + e^{i(w_{h}t - \bm{k}\cdot\bm{x} +\alpha)} \right) \ , \label{cross} & 
\end{numcases}
where $\omega_{h}$ is an angular frequency of the gravitational wave and $\alpha$ represents 
a difference of the phases of polarizations.
Note that the polarization tensors  can be explicitly constructed as
\begin{eqnarray}
  e _{ij}^{(+)} &=& \frac{1}{\sqrt{2}}\left(
    \begin{array}{ccc}
      \cos\theta^{2} & 0 & -\cos\theta \sin\theta \\
      0 & -1 & 0 \\
      -\cos\theta \sin\theta & 0 & \sin\theta^{2}
    \end{array} 
  \right) , \label{lipo1}  \\
  e _{ij}^{(\times)} &=& \frac{1}{\sqrt{2}}\left(
    \begin{array}{ccc}
      0 & \cos\theta & 0 \\
      \cos\theta & 0 & -\sin\theta \\
      0 & -\sin\theta & 0
    \end{array} 
  \right) .  \label{lipo2}
\end{eqnarray}
In the above Eqs.\,(\ref{lipo1}) and (\ref{lipo2}), we defined the $+$ mode as a deformation in the $y$-direction.

Then substituting Eqs.\,(\ref{expan})-(\ref{lipo2}) into 
the total Hamiltonian (\ref{toHami}),
moving on to the Fourier space and using the rotating wave approximation,
one can deduce
\begin{equation}
   H_{{\rm tot}} \simeq 2\mu_{B} B_{z}  \hat{c}^{\dagger} \hat{c} +  
 g_{eff} \left(    \hat{c}^{\dagger} e^{-i\omega_{h}t} +    \hat{c} e^{i\omega_{h}t} \right)  ,
                      \label{dri}
\end{equation}
where $\hat{c} = \hat{c}_{k=0}$ and
\begin{equation}
g_{eff} = \frac{\sqrt{2}\pi^{2}}{60}
\left( \frac{l}{\lambda} \right)^{2} \mu_{B} B_{z} \sin\theta \sqrt{N} 
\left[ \cos^{2} \theta \, (h^{(+)})^{2} + (h^{(\times)})^{2} + 2 \cos\theta\sin\alpha \, h^{(+)}h^{(\times)} 
\right]^{1/2}  \ , \label{effco}
\end{equation}
is an effective coupling constant between the gravitational waves and the magnons.
The parameters $l$ and $\lambda = 2\pi / \omega_{h}$ are 
the radius of the (spherical) ferromagnetic sample 
and the wavelength of the gravitational wave.
We note that the sum over the spin sites $i$ was evaluated as 
\begin{equation}
  \sum_{i} xx =  \sum_{i} yy = \sum_{i} zz \simeq 
    \frac{1}{L^{3}} \iiint^{l}_{0} r^{2} \sin\zeta   
    \left( r \cos\zeta \right)^{2} dr d\zeta d\phi
    = \frac{4\pi}{15} \frac{l^{5}}{L^{3}}  \ ,  \label{integration}
\end{equation}
where $L$ is a lattice constant, which is related to the number of spins as 
$N = \left( \frac{4\pi}{3}l^{3} \right) / L^{3}$.

From Eq.\,(\ref{effco}), we see that the effective coupling constant has gotten a huge factor $\sqrt{N}$.
Moreover, in order to obtain a coordinate-independent expression of $g_{eff}$,
it is useful to use the Stokes parameters:
\begin{equation}
g_{eff} = \frac{\sqrt{2}\pi^{2}}{60} 
          \left( \frac{l}{\lambda} \right)^{2} 
          \mu_{B} B_{z} \sin\theta \sqrt{N} 
          \left[ \frac{1+\cos^{2}\theta}{2} \, I - 
          \frac{\sin^{2}\theta}{2} \, Q  + \cos \theta \, V  \right]^{1/2}  \ , \label{effco2}
\end{equation}
where the Stokes parameters are defined by 
\begin{eqnarray}
  \begin{cases}
   I = (h^{(+)})^{2} + (h^{(\times)})^{2} \ , & \\
   Q = (h^{(+)})^{2} - (h^{(\times)})^{2}  \ , & \\
   U = 2 \cos\alpha \, h^{(+)} h^{(\times)}  \ , & \\
   V = 2 \sin\alpha \, h^{(+)} h^{(\times)}  \ . &
  \end{cases} \label{stokes}
\end{eqnarray}
They satisfy $I^{2} = U^{2} + Q^{2} +V^{2}$.
We see that the effective coupling constant depends on the polarizations.
Note that the stokes parameters $Q$ and $U$ transform as 
\begin{equation}
\binom{Q'}{U'} = \begin{pmatrix}
\cos 4\psi & \sin 4\psi \\
-\sin 4\psi & \cos 4\psi   \end{pmatrix}
\binom{Q}{U}
\end{equation}
where $\psi$ is the rotation angle around $\bm{k}$.

The second term in Eq.\,(\ref{dri}) shows that 
planar gravitational waves induce the resonant spin precessions and/or 
the excitation of magnons
if the angular frequency of the gravitational waves is near the 
Lamor frequency, $2\mu_{B} B_{z}$.
It is worth noting that the situation is similar to the resonant bar experiments~\cite{Maggiore:1999vm} where 
planar gravitational waves excite phonons in a bar detector.
%
%
%
%
%
%
%
%
%
%
%
%
%

Let us show the ability of magnon gravitational detectors by giving constraints 
on high frequency gravitational waves.
Recently, measurements of resonance fluorescence of magnons induced by the axion dark matter 
was conducted and upper bounds on an axion-electron coupling constant have been 
obtained~\cite{Crescini:2018qrz,Flower:2018qgb}.
Such an axion-magnon resonance~\cite{Barbieri:1985cp}
has a similar mechanism to our graviton-magnon resonance.
Therefore, we can utilize these experimental results to give the upper bounds on the amplitude of GHz gravitational waves~\cite{Ito:2019wcb}.

The interaction hamiltonian which describe the axion-magnon resonance is given by
\begin{equation}
  \mathcal{H}_{a} 
  = \tilde{g}_{eff} \left(    \hat{c}^{\dagger} e^{-im_{a}t} +    \hat{c} e^{im_{a}t} \right) \ , \label{axion}
\end{equation}
where $\tilde{g}_{eff}$ is an effective coupling constant between an axion and a magnon.
Notice that the axion oscillates with a frequency determined by the axion mass $m_{a}$.
One can see that this form is the same as the interaction term in Eq.\,(\ref{dri}).
Through the hamiltonian (\ref{axion}), 
$\tilde{g}_{eff}$ is related to an axion-electron coupling constant in~\cite{Crescini:2018qrz,Flower:2018qgb}.
Then the axion-electron coupling constant can be converted to 
$\tilde{g}_{eff}$ by using parameters, such as the energy density of the axion dark matter, which 
are explicitly given in~\cite{Crescini:2018qrz,Flower:2018qgb}.
Therefore constraints on $\tilde{g}_{eff}$ (95\% C.L.) can be read from the 
constraints on the axion-electron coupling constant given in~\cite{Crescini:2018qrz} and \cite{Flower:2018qgb},
respectively, as follows:
\begin{eqnarray}
  \tilde{g}_{eff} < 
  \begin{cases}
    3.5 \times 10^{-12} \ {\rm eV}  \ , & \\
    3.1 \times 10^{-11} \ {\rm eV}  \ . &  
  \end{cases} \label{aco}
\end{eqnarray}

\begin{figure}[H]
\centering
\includegraphics[width=10cm]{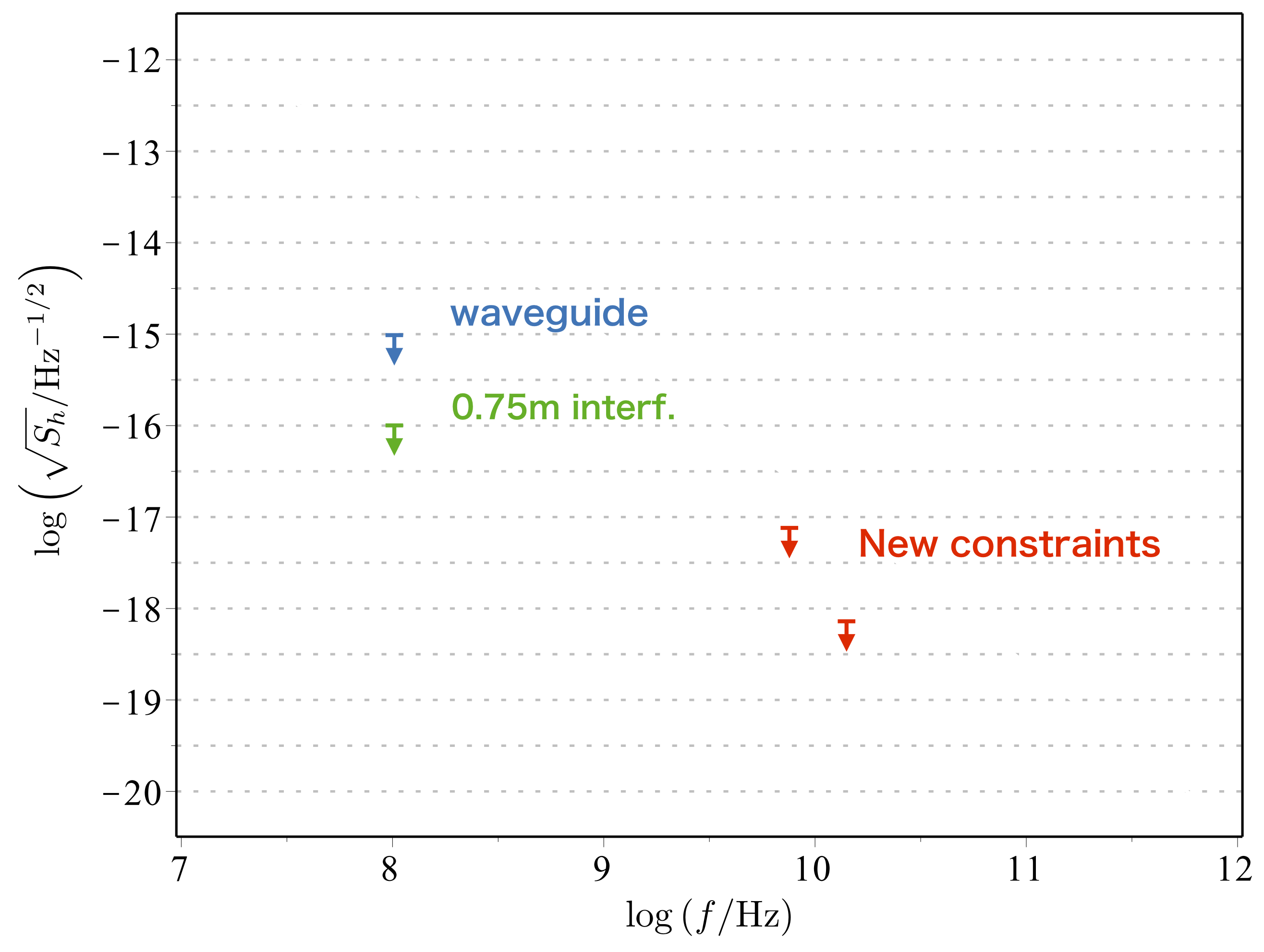}
\caption{Several experimental sensitivities and constraints on high frequency gravitational waves are depicted.
The blue color represents an upper limit on stochastic gravitational waves
by waveguide experiment using an interaction between 
electromagnetic fields and gravitational waves~\cite{0264-9381-23-22-007}.
The green one is the upper limit on stochastic gravitational waves,
obtained by the 0.75 m interferometer~\cite{Akutsu:2008qv}.
Our new constraints on continuous gravitational waves are plotted with a red color, which also 
represent the sensitivity of the magnon gravitational wave detector for stochastic gravitational waves.
}
\label{GWfig}
\end{figure}

It is easy to convert the above constraints 
to those on the amplitude of gravitational waves appearing in the effective coupling constant (\ref{effco2}).
Indeed,  we can read off the external magnetic field $B_{z}$ and the number of electrons $N$ 
as $(B_{z},N) = (0.5 \,{\rm T}, \ 5.6\times 10^{19} )$ from~\cite{Crescini:2018qrz} and 
$(B_{z},N) = (0.3 \,{\rm T}, \ 9.2\times 10^{19} )$ from~\cite{Flower:2018qgb}, respectively.
The external magnetic field $B_{z}$ determines the frequency of gravitational waves we can detect.
Therefore, using Eqs.\,(\ref{effco2}), (\ref{aco}) and the above parameters, one can put upper limits on  
gravitational waves at frequencies determined by $B_{z}$.
Since \cite{Crescini:2018qrz} and \cite{Flower:2018qgb} focused on the direction of Cygnus and set the 
external magnetic field to be perpendicular to it, 
we probe continuous gravitational waves coming from Cygnus with $\theta = \frac{\pi}{2}$
(more precisely, $\sin\theta=0.9$ in \cite{Flower:2018qgb}).
We also assume no linear and circular polarizations, i.e.,  $Q'=U'=V=0$.
Consequently, experimental data \cite{Crescini:2018qrz} and \cite{Flower:2018qgb} enable us to constrain
the characteristic amplitude of gravitational waves defined by 
$h_{c} = h^{(+)} = h^{(\times)}$ as
\begin{eqnarray}
  h_{c} \sim
  \begin{cases}
    1.3 \times 10^{-13}  \quad {\rm at} \    14  \  {\rm GHz} \ , & \\
    1.1 \times 10^{-12}  \quad {\rm at} \   8.2 \  {\rm GHz}  \ , &
  \end{cases} 
\end{eqnarray}
at $95$ \% C.L., respectively.
In terms of the spectral density defined by $S_{h} = h_{c}^{2}/2f$ and 
the energy density parameter defined by $\Omega_{GW} = 2\pi^{2} f^{2} h_{c}^{2} / 3 H_{0}^{2}$ ($H_{0}$ is 
the Hubble parameter),
the upper limits at $95$ \% C.L. are 
\begin{eqnarray}
  \sqrt{S_{h}} \sim
  \begin{cases}
    7.5 \times 10^{-19}  \ [{\rm Hz}^{-1/2}]  \quad {\rm at} \   14  \  {\rm GHz} \ , & \\
    8.7 \times 10^{-18}  \ [{\rm Hz}^{-1/2}]  \quad {\rm at} \   8.2 \  {\rm GHz}  \ , &
  \end{cases}   \label{Cons}
\end{eqnarray}
and 
\begin{eqnarray}
  h_{0}^{2} \Omega_{GW} \sim
  \begin{cases}
    2.1 \times 10^{29}   \quad {\rm at} \   14  \  {\rm GHz} \ , & \\
    5.5 \times 10^{30}   \quad {\rm at} \   8.2 \  {\rm GHz}  \ . &
  \end{cases}   \label{aaaaa}
\end{eqnarray}
We depict the limits on the spectral density with several other gravitational wave experiments 
in Fig.\,\ref{GWfig}.
%
%
%
%
%
%
%
%
%
%
%
%
%
%
%
%
%
%
%
%
%
%
%
%
%
%
\section{Conclusion}
In order to detect high frequency gravitational waves, we developed a new detection method. 
Using Fermi normal coordinates and taking the non-relativistic limit,
we obtained the Hamiltonian for non-relativistic fermions in Fermi normal coordinates
for general curved spacetime.
This Hamiltonian is applicable for any curved spacetime background as long as 
one can treat a curvature perturbatively.
Therefore, our formalism is useful to consider gravitational effects on non-relativistic fermions,
which is usual  in  condensed matter systems.

In the section \ref{GMreso},
we focused on the interaction between a spin of a fermion and gravitational waves, expressed by 
the third line in Eq.\,(\ref{Hddd2}).
It turned out that gravitational waves can excite magnons.
Moreover, we explicitly demonstrated how to use magnons for detecting high frequency gravitational waves 
and gave upper limits on
the spectral density of continuous gravitational waves ($95$ \% C.L.): 
$7.5 \times 10^{-19}  \  [{\rm Hz}^{-1/2}]$ at 14 GHz and
$8.7 \times 10^{-18}  \  [{\rm Hz}^{-1/2}]$ at 8.2 GHz, respectively,
by utilizing results of magnon experiments.
Interestingly, there are several theoretical models predicting high frequency gravitational waves
which are within the scope of our method~\cite{Kuroda:2015owv}.

The graviton-magnon resonance is also useful for probing stochastic gravitational waves with
almost the same sensitivity illustrated in Fig.\,\ref{GWfig}.
Although the current  sensitivity is still not sufficient for putting a meaningful constraint on 
stochastic gravitational waves, 
it is important to pursue the high frequency stochastic gravitational wave search
for future gravitational wave physics.
Moreover, we can probe burst gravitational waves of any wave form
if the duration time is smaller than the relaxation time of a system.
The situation is the same as for resonant bar detectors~\cite{Maggiore:1900zz,Astone:2010mr}.
For instance, in the measurements~\cite{Crescini:2018qrz,Flower:2018qgb}, 
the relaxation time is about $0.1$ $\upmu$s which is
determined by the line width of the ferromagnetic sample and the cavity.
If the duration of a burst of gravitational waves is smaller than $0.1$ $\upmu$s, 
we can detect it.
Furthermore, improving the line width of the sample and the cavity not only leads to 
detecting burst gravitational waves but also to increasing the sensitivity.
As another way to improve the sensitivity of the magnon gravitational wave detector,
quantum nondemolition measurement may be promising~\cite{Tabuchi405,TABUCHI2016729,Lachance-Quirione1603150}.
\begin{acknowledgments}
A.\,I. was supported by Grant-in-Aid for JSPS Research Fellow and JSPS KAKENHI Grant No.JP17J00216.
J.\,S. was in part supported by JSPS KAKENHI Grant Numbers JP17H02894, JP17K18778.
This research was supported by the Munich Institute for Astro - and Particle Physics (MIAPP) which is funded by the Deutsche Forschungsgemeinschaft (DFG,German Research Foundation) under Germany's Excellence Strategy-EXC-2094-390783311.
\end{acknowledgments}
\appendix
\section{Fermi normal coordinates}\label{Fermi}
One can construct local inertial coordinates along a geodesic of a particle, 
the so-called Fermi normal coordinates~\cite{Manasse:1963zz}.
An observer on the earth is freely falling when gravity of the
earth, which will be taken into account in the Appendix \ref{EarthGra}, is negligible.
Thus, the Fermi normal coordinates describe the frame
 used in an experiment.
In this appendix, we briefly review how to construct the Fermi normal coordinates~\cite{Manasse:1963zz}.

\begin{figure}[H]
\centering
\includegraphics[width=6.5cm]{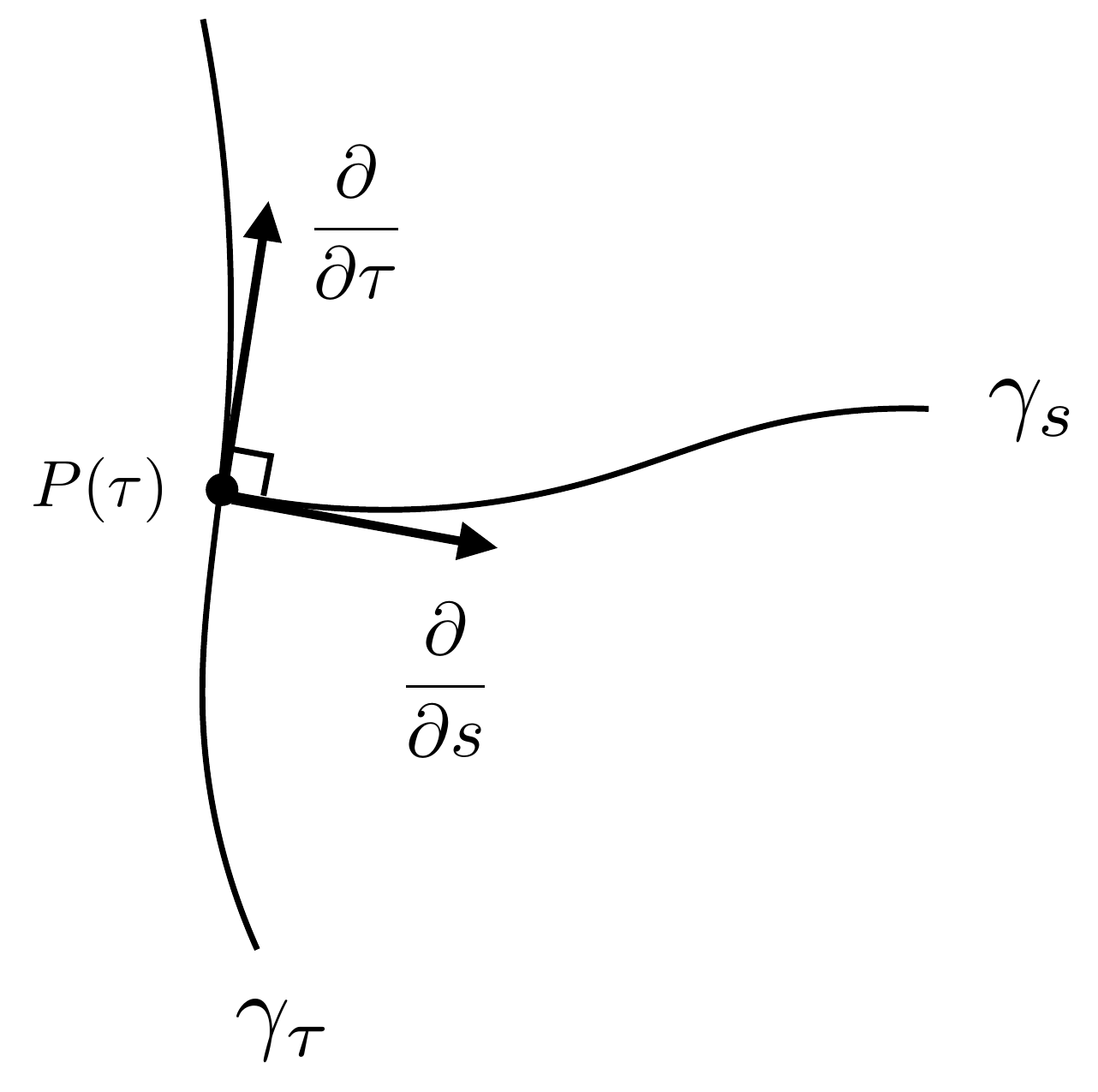} 
\caption{A timelike geodesic $\gamma_{\tau}$ is parametrized by a proper time $\tau$ and 
a spacelike geodesic $\gamma_{s}$ is parametrized by a proper distance $s$, which
is orthogonal to $\gamma_{\tau}$ at the crossing point.} \label{Fer}
\end{figure}

We consider a timelike geodesic $\gamma_{\tau}$ parametrized by a proper time $\tau$ and 
specify a point on the geodesic by $P(\tau)$. 
We also consider a spacelike geodesic $\gamma_{s}$ orthogonal to 
$\gamma_{\tau}$ at $P(\tau)$,
which is parametrized by a proper distance $s$.%
\footnote{
One can use an affine parameter instead of $s$, which does not change the following discussion. 
}
We set  the crossing point as $s=0$. 
The situation is illustrated in Fig.\,\ref{Fer}.
\noindent
Then, the Fermi normal coordinates which are locally inertial frames along $\gamma_{\tau}$ 
are defined as follows:
\begin{equation}
  x^{0} = \tau , \quad x^{i} = \alpha^{i} s \ .  \label{Feco}
\end{equation}
The bases of the Fermi normal coordinates, $\frac{\partial}{\partial x^{\mu}}$,
are parallelly transported along the geodesic $\gamma_{\tau}$ and 
$\alpha^{i}$ are the components of the tangent vector $\frac{\partial}{\partial s}$, namely,
\begin{equation}
  \frac{\partial}{\partial s} = \alpha^{i} \frac{\partial}{\partial x^{i}}  \ .
\end{equation}
Also, the bases, $\frac{\partial}{\partial x^{\mu}}$, are taken to be orthonormal
by utilizing the degree of rescaling $\alpha^{i}$.
Thus, the metric in the Fermi normal coordinates is given by $\eta_{\mu\nu}$ on the geodesic $\gamma_{\tau}$.\footnote{
Note that orthonormality holds at any point on the geodesic
 $\gamma_{\tau}$ if 
it is satisfied at one point on $\gamma_{\tau}$, because the parallel transport 
keeps orthonormality.
}

Let us show that the Fermi normal coordinates (\ref{Feco}) 
 are indeed local inertial frames, namely,
the Christoffel symbols are zero along the geodesic $\gamma_{\tau}$.
First, because the bases of Fermi normal coordinates are 
parallelly transformed along $\gamma_{\tau}$, we have
\begin{eqnarray}
  0 &=& \left( \frac{\partial}{\partial x^{\nu}} \right)^{\mu}_{;\alpha}  
  \left( \frac{\partial}{\partial \tau} \right)^{\alpha}   \nonumber \\
  &=&  \left( \delta^{\mu}_{\nu} \right)_{;\alpha} \delta^{\alpha}_{0}  \nonumber \\
  &=&  \Gamma^{\mu}_{\nu 0}(x^{0}=\tau ,x^{i}=0)   \nonumber \\
  &=&  \Gamma^{\mu}_{\nu 0}|_{\gamma_{\tau}}       \ ,  \label{chri0}
\end{eqnarray}
where we used the fact that the vector components of the bases in the Fermi normal coordinates are
$\left( \frac{\partial}{\partial x^{\nu}} \right)^{\mu} = \delta^{\mu}_{\nu}$.
On the other hand,
on the spacelike geodesic $\gamma_{s}$, 
the geodesic equation 
\begin{equation}
  \frac{d^{2} x^{\mu}}{ds^{2}} 
   + \Gamma^{\mu}_{\alpha\beta} \frac{dx^{\alpha}}{ds} \frac{dx^{\beta}}{ds}  = 0  \ , \label{gams}
\end{equation}
is satisfied.
Using (\ref{Feco}) in Eq.\,(\ref{gams}), we obtain
\begin{equation}
  \Gamma^{\mu}_{ij}(x^{0}=\tau, x^{i}=\alpha^{i}s) \, \alpha^{i} \alpha^{j} = 0 \ .
\end{equation}
In particular on $\gamma_{\tau}$, namely at $s=0$,
we conclude that 
\begin{equation}
  \Gamma^{\mu}_{ij}(x^{0}=\tau, x^{i}=0) =  \Gamma^{\mu}_{ij}|_{\gamma_{\tau}} =  0 \ .  \label{chrii}
\end{equation}
Therefore, from Eqs.\,(\ref{chri0}) and (\ref{chrii}), we see that the Christoffel symbols vanish 
along the timelike geodesic $\gamma_{\tau}$.

Now one can calculate the metric components in the vicinity of the geodesic
$\gamma_{\tau}$ in the Fermi normal coordinates.
In a situation that a curvature scale is much larger than that of 
a system we treat, we can expand the metric in terms of the coordinates $x^{\mu}$.
The first order term vanishes by definition. For our purpose,
it is enough to calculate the metric up to the second order.

Note that the Christoffel symbols vanish along the geodesic $\gamma_{\tau}$
\begin{equation}
  \Gamma^{\mu}_{\nu\lambda,0}|_{\gamma_{\tau}} = 0 \ .   \label{ggaa}
\end{equation}
Then, from the definition of the Riemann tensor, we find
\begin{equation}
  \Gamma^{\mu}_{\nu0,\lambda}|_{\gamma_{\tau}} = R^{\mu}{}_{\nu\lambda 0}|_{\gamma_{\tau}} \ .  \label{gaderi}
\end{equation}
To go further, we use the geodesic deviation equation:
\begin{eqnarray}
  \frac{d^{2} \xi^{\mu}}{d \lambda ^{2}} 
    + 2 \frac{d\xi^{\alpha}}{d\lambda} \Gamma^{\mu}_{\alpha\beta} u^{\beta} 
    + \left(  
    R^{\mu}{}_{\alpha\gamma\beta}
    + \Gamma^{\mu}_{\alpha\gamma,\beta}
    + \Gamma^{\mu}_{\beta\delta} \Gamma^{\delta}_{\alpha\gamma} 
    - \Gamma^{\mu}_{\gamma\delta} \Gamma^{\delta}_{\alpha\beta}   
    \right) u^{\alpha} u^{\beta} \xi^{\gamma}  = 0  \ ,  \label{geodevieq2}
\end{eqnarray}
where $\lambda$ can be either $\tau$ or $s$.
We notice that a point on $\gamma_{s}$ is specified by the parameters $(\tau, s, \alpha^{i})$.
Then, as to the spacelike geodesic $\gamma_{s}$, one can consider two deviation vectors; 
one is $\left( \frac{\partial}{\partial \tau} \right)_{s,\alpha^{i}}$ and
the other is $\left( \frac{\partial}{\partial \alpha^{i}} \right)_{\tau,s}$.
The vector $\left( \frac{\partial}{\partial \tau} \right)_{s,\alpha^{i}}$ represents a deviation 
between two spacelike geodesics which stem from different points on $\gamma_{\tau}$ and
$\left( \frac{\partial}{\partial \alpha^{i}} \right)_{\tau,s}$ represents a deviation 
between two spacelike geodesics which stem from the same point $P(\tau)$ on $\gamma_{\tau}$. 
Substituting 
$\xi^{\mu} 
= \left( \frac{\partial}{\partial \tau} \right)^{\mu}_{s,\alpha^{i}} 
= \delta^{\mu}_{0}$ into 
Eq.\,(\ref{geodevieq2}) yields 
\begin{equation}
 \left( \Gamma^{\mu}_{i0,j}|_{\gamma_{\tau}} - R^{\mu}{}_{ij0}|_{\gamma_{\tau}} \right) 
   \alpha^{j} \alpha^{k}  =  0   \ ,
\end{equation}
which is automatically satisfied because of Eq.\,(\ref{gaderi}).
On the other hand, 
substituting 
$\xi^{\mu} 
= \left( \frac{\partial}{\partial \alpha^{i}} \right)_{\tau,s}
= s \delta^{\mu}_{i}$ into
Eq.\,(\ref{geodevieq2}), we obtain
\begin{equation}
  2 \Gamma^{\mu}_{ij} \alpha^{j} 
  + s R^{\mu}{}_{jik}|_{\gamma_{\tau}} \alpha^{j} \alpha^{k} 
  + s \Gamma^{\mu}_{ij,k}|_{\gamma_{\tau}} \alpha^{j} \alpha^{k} 
  + \mathcal{O}(s^{2}) 
  = 0  \ .   \label{ddee}
\end{equation}
The first term in Eq.\,(\ref{ddee}) can be expanded in powers of $s$ as 
\begin{eqnarray}
  2 \Gamma^{\mu}_{ij} \alpha^{j} 
  &=&  2 \Gamma^{\mu}_{ij}|_{\gamma_{\tau}}   \alpha^{j}
     + 2 s \left( \frac{\partial}{\partial s} \Gamma^{\mu}_{ij} 
                  \right)_{{\rm at} \, \gamma_{\tau}}
                 \alpha^{j}  \nonumber \\
  &=&  2 s \Gamma^{\mu}_{ij,k}|_{\gamma_{\tau}} \alpha^{j} \alpha^{k}   \ .  \label{glaex}
\end{eqnarray}
From Eqs.\,(\ref{ddee}) and (\ref{glaex}), we find a relation 
\begin{equation}
   \left( \Gamma^{\mu}_{ij,k}|_{\gamma_{\tau}}  
   + \frac{1}{3} R^{\mu}{}_{jik}|_{\gamma_{\tau}} 
   \right)  \alpha^{j} \alpha^{k} = 0  \ .
\end{equation}
It implies that the symmetric part of the indices $j$ and $k$ in the parenthesis should be zero, i.e.,
\begin{equation}
  \Gamma^{\mu}_{ij,k}|_{\gamma_{\tau}} + \Gamma^{\mu}_{ik,j}|_{\gamma_{\tau}}
  = - \frac{1}{3}\left( R^{\mu}{}_{jik}|_{\gamma_{\tau}} + R^{\mu}{}_{kij}|_{\gamma_{\tau}} \right) \ .
\end{equation}
After  a little algebra, this can be solved  as
\begin{equation}
  \Gamma^{\mu}_{ij,k}|_{\gamma_{\tau}} =
    - \frac{1}{3}\left( R^{\mu}{}_{ijk}|_{\gamma_{\tau}} + R^{\mu}{}_{jik}|_{\gamma_{\tau}} \right) \ . 
    \label{Gijk}
\end{equation}

From the definition of the Christoffel symbol, we have
\begin{equation}
  g_{\mu\nu,\lambda} =  g_{\mu\alpha} \Gamma^{\alpha}_{\nu\lambda} 
                      + g_{\nu\alpha} \Gamma^{\alpha}_{\mu\lambda} \ .
\end{equation}
Differentiating it with respect to $x^{\sigma}$ leads to
\begin{equation}
  g_{\mu\nu,\lambda\sigma}|_{\gamma_{\tau}} = 
      \eta_{\mu\alpha} \Gamma^{\alpha}_{\nu\lambda ,\sigma}|_{\gamma_{\tau}}
    + \eta_{\nu\alpha} \Gamma^{\alpha}_{\mu\lambda ,\sigma}|_{\gamma_{\tau}} \ .  \label{metmet}
\end{equation}
Using Eqs.\,(\ref{ggaa}), (\ref{gaderi}) and (\ref{Gijk}) in Eq.\,(\ref{metmet}),
one can deduce  $g_{\mu\nu,0\lambda} = 0$ and the following:
\begin{eqnarray}
  g_{00,ij} &=& -2 R_{0i0j}|_{\gamma_{\tau}} \ , \nonumber \\
  g_{0i,jk} &=& -\frac{2}{3} 
  \left( R_{0jik}|_{\gamma_{\tau}}  + R_{0kij}|_{\gamma_{\tau}} \right)   \ ,  \nonumber \\
  g_{ij,kl} &=& - \frac{1}{3}
  \left( R_{ikjl}|_{\gamma_{\tau}} + R_{iljk}|_{\gamma_{\tau}} \right)  \ .
\end{eqnarray}
Thus, up to the quadratic order of the coordinates, 
the metric components in the Fermi normal coordinates are given by
\begin{eqnarray}
  g_{00} &=& - 1 - R_{0i0j}|_{\gamma_{\tau}} x^{i} x^{j} \ ,  \label{met01} \\
  g_{0i} &=& -\frac{2}{3} R_{0jik}|_{\gamma_{\tau}} x^{j} x^{k} \ ,  \label{met02} \\
  g_{ij} &=& \delta_{ij} - \frac{1}{3} R_{ikjl}|_{\gamma_{\tau}} x^{k} x^{l} \ .  \label{met03}
\end{eqnarray}
We note that the Riemann tensor is evaluated on the timelike geodesic $\gamma_{\tau}$, hence it only depends on $x^{0}$.
It should be mentioned that the Riemann tensor in Eqs.\,(\ref{met01})-(\ref{met03}) is calculated in the Fermi normal coordinates.
However, the Riemann tensor for the linear perturbations around the flat spacetime background is invariant under gauge transformations.
In this case, the Riemann tensor constructed in the Fermi normal coordinates is the same as that in the transverse traceless gauge.
Therefore, one can use (\ref{Rie2}) in Eqs.\,(\ref{met01})-(\ref{met03})
when we consider gravitational waves on the flat spacetime background.
\section{Proper detector frame}\label{EarthGra}
In the Appendix \ref{Fermi}, we constructed local inertial coordinates along a geodesic for
a freely falling observer, namely the Fermi normal coordinates.
However, an observer bound on the earth is not freely falling.
This is because the observer accelerates against the center of the earth 
with $g = 9.8 \  {\rm m}/{\rm s}^{2}$
and has the rotational motion since the earth is rotating.
In this appendix, we take into account these effects of the earth~\cite{Misner:1974qy,Ni:1978zz}.
We will see that these effects are negligible in the discussion in the text.

The procedure 
is almost the same as the case of the Fermi normal coordinates.
We first consider a timelike curve $\gamma_{\tau}$ parametrized by 
 $\tau$ and  
construct a spacelike geodesic $\gamma_{s}$ parametrized by a proper distance $s$, which is orthogonal to the geodesic $\gamma_{\tau}$ at $s=0$.
The situation is illustrated in Fig.\,\ref{Fer}.
The difference from the Fermi normal coordinates
appears in the transportation of the orthonormal bases $\bm{e}_{\mu}$ which 
cover small region around a point on the curve $\gamma_{\tau}$.
The bases $\bm{e}_{\mu}$ are parallelly transported
 along $\gamma_{\tau}$,
i.e. $\frac{d}{d\tau} \bm{e}_{\mu} = 0$, in the construction of the Fermi normal coordinates%
\footnote{
Here, the coordinate bases are not restricted to those given by Eq.\,(\ref{Feco}).}%
,
while $e_{\mu}$ in the present case are transported as follows~\cite{Misner:1974qy}:
\begin{eqnarray}
  \frac{D}{d\tau} \bm{e}_{\mu} 
   &=& - \bm{\Omega} \cdot \bm{e}_{\alpha}  \nonumber \\
   &=&  \Omega^{\mu\nu} \left( \bm{e}_{\mu} \wedge \bm{e}_{\nu} \right) \cdot \bm{e}_{\alpha} \nonumber \\
   &=&  - \bm{e}_{\alpha} \Omega^{\alpha}_{\ \mu} \ , \label{tetevo}
\end{eqnarray}
where $\Omega^{\mu\nu}$ is an infinitesimal Lorentz transformation defined by
\begin{eqnarray}
  \Omega^{\mu\nu} &=& \left( a^{\mu}u^{\nu} - a^{\nu} u^{\mu} \right)
                      + u_{\alpha} \omega_{\beta} \epsilon^{\alpha\beta\mu\nu}  \nonumber \\
                  &=& _{({\rm F})}\Omega^{\mu\nu}  +  _{({\rm R})}\Omega^{\mu\nu} \ . \label{Omega}
\end{eqnarray}
Here, we defined the four velocity
\begin{equation}
  u^{\mu} = \frac{d x^{\mu}}{d \tau} \ ,
\end{equation}
 the four acceleration
\begin{equation}
  a^{\mu} = \frac{d u^{\mu}}{d \tau} \ , \label{accel}
\end{equation}
 and 
$\omega_{\mu}$ represents an angular velocity of rotation of spatial bases $\bm{e}_{i}$.
Note that the orthonormality of the bases holds under the evolution (\ref{tetevo})
as a consequence of the anti-symmetricity of $\Omega^{\mu\nu}$.

One can see that
$_{({\rm R})}\Omega^{\mu\nu}$ 
represents just a three dimensional rotation 
in a four dimensional covariant form.
In fact, in the rest frame, i.e. $u^{\mu} = (1, 0,0,0)$, we obtain
\begin{eqnarray}
  - \bm{e}_{\alpha} \, _{({\rm R})}\Omega^{\alpha}_{\mu} 
      &=&  - \bm{e}_{\alpha} u_{\gamma} \omega_{\beta} \epsilon^{\gamma\beta\alpha}_{\ \ \ \ \mu} \nonumber \\
      &=&  \omega_{i} \bm{e}_{j} \epsilon^{0 i j}_{\ \ \ \mu}  \nonumber \\
      &=&  \left( \bm{\omega} \times \bm{e}_{j} \right)_{\mu=k} \ ,
\end{eqnarray}
where we identified the label of the bases $\bm{e}_{\mu}$ as the component of them due to 
orthonormality to obtain the last equality and 
$\mu=k$ represents the fact that $\mu$ takes a spatial index. 
For the observer on the earth,
$\bm{\omega}$ is the angular velocity of the earth.

The transformation $_{{\rm (F)}}\Omega^{\mu\nu}$ is
called the Fermi-Walker transport.
Consider an accelerating observer with magnitude of the gravity of the earth,
$a^{\mu}a_{\mu} = g^{2}$, along $x^{1}$-coordinate in an inertial frame%
\footnote{
In the rest frame for the observer, 
the Newton equation
$\frac{d^{2} x^{i}}{(d x^{0})^{2}} - g = 0$
holds. Here, we used the fact that the 0-component of $a^{\mu}$ is zero because
$a^{\mu}$ is orthogonal to $u^{\mu}$ 
and $u^{\mu} = \delta^{\mu}_{0}$ in the rest frame. 
Therefore, the relation of the relativistically invariant quantity, $a^{\mu} a_{\mu} = g$, is
satisfied as in Newtonian gravity.
}.
Then, because an acceleration vector defined by (\ref{accel}) is orthogonal to the four velocity, we have
\begin{equation}
  a^{\mu} u_{\mu} =  - a^{0} u^{0} + a^{i} u^{i} = 0 \ . \label{a}
\end{equation}
Using the above and an explicit relation
\begin{equation}
  a^{\mu} a_{\mu} = - a^{0}a^{0} + a^{1}a^{1} = g^{2} \ ,
\end{equation}
we can obtain the following equations:
\begin{eqnarray}
\left\{
\begin{array}{l}
    a^{0} = \frac{u^{0}}{d\tau} = g u^{1} \ , \\
    a^{1} = \frac{u^{1}}{d\tau} = g u^{0} \ .   \label{renritsu}
\end{array}
\right.
\end{eqnarray}
A solution of Eqs.\,(\ref{renritsu}) is given by
\begin{eqnarray}
\left\{
\begin{array}{l}
    t = g^{-1} \sinh \left( g \tau \right) \ , \\
    x^{1} = g^{-1} \cosh \left( g \tau \right) \ ,   \label{renritsusol}
\end{array}
\right.
\end{eqnarray}
which is a hyperbolic world curve, indeed, $x^{2} -t^{2} = g^{-2}$.
The hyperbolic curve is invariant under a Lorentz boost from the inertial coordinate $(t, x^{1})$ to another one.
Since $\tau$ dependence appears in Eqs.\,(\ref{renritsusol}), 
one can construct the rest frame for the accelerating observer 
at instant $\tau$ by 
doing a Lorentz boost transformation depending on $\tau$.
Such a Lorentz boost, which is a four dimensional rotation of a plane 
spanned by $u^{\mu}$ and $a^{\mu}$, would be expressed by $_{{\rm (F)}}\Omega^{\mu\nu}$.
Indeed, for an observer accelerating along the $x^{1}$-direction, 
we have
\begin{equation}
  _{{\rm (F)}}\Omega^{0x^{1}} = -g \ , 
\end{equation}
and the other components of $_{{\rm (F)}}\Omega^{\mu\nu}$ vanish.
Thus, the four vector $x^{\mu} = (\tau, 0, 0, 0)$, 
after the infinitesimal Lorentz transformation conducted by $_{{\rm (F)}}\Omega^{\mu\nu}$, 
is given by
\begin{eqnarray}
  d \left( x^{0'} - x^{0} \right) &=& x^{\mu} \,  _{{\rm (F)}}\Omega_{\mu 0'} d\tau  \nonumber \\
         &=& 0 \ .
\end{eqnarray}
Hence, we get 
\begin{equation}
  \frac{d x^{0'}}{d \tau} = 0 \ .
\end{equation}
This is consistent with the first equation in (\ref{renritsusol}) when
$g\tau \ll 1$.
Similarly, we obtain 
\begin{eqnarray}
  d \left( x^{1'} - x^{1} \right) &=& x^{\mu} \, _{{\rm (F)}}\Omega_{\mu 1'} d\tau  \nonumber \\
         &=& \tau g d\tau \ ,
\end{eqnarray}
which leads to
\begin{equation}
  \frac{ d x^{1'}}{d \tau} =  g \tau \ .
\end{equation}
This is consistent with the second equation in (\ref{renritsusol}) when
$g\tau \ll 1$.
Therefore,
we find that $_{{\rm (F)}}\Omega^{\mu\nu}$ correctly represents an infinitesimal Lorentz transformation
which connects a rest frame to an accelerating frame relative to the rest frame.
Now, we can understand the meaning of the Fermi-Walker transport in Eq.\,(\ref{tetevo}).
At one point on $\gamma_{\tau}$, one can construct a rest frame for an accelerating observer,
but after a certain duration the frame is not a rest frame for the observer anymore.
In order to keep a frame as a rest frame at any time $\tau$, the bases of the frame should be developed by the Fermi-Walker transport.
Thus, we obtain a coordinate system moving with an accelerating observer.

From now on, we use coordinate bases specified by Eq.\,(\ref{Feco}):
\begin{equation}
  x^{0} = \tau , \quad x^{i} = \alpha^{i} s \ ,  \label{Feco2}
\end{equation}
and 
get an explicit expression for the metric in the proper detector coordinate which 
is moving with an accelerating observer due to the earth. 
The procedure is similar to the case of the Fermi normal coordinates in the Appendix A. 

From Eq.\,(\ref{tetevo}), we obtain a relation:
\begin{equation}
  \Gamma^{\alpha}_{\mu 0} =  \Omega^{\alpha}_{\ \mu} \ . \label{GOrela}
\end{equation}
Using 
$u^{\mu} = (1,0,0,0)$ and $a^{\mu} = (0, a^{i})$ in the definition  (\ref{Omega}),
we have
\begin{equation}
  \Omega^{0}_{\ i} = a_{i}, \quad \Omega^{i}_{\ j} = - \epsilon^{0ijk}\omega_{k} \ .  \label{o01oij}
\end{equation}
Thus together with Eqs.\,(\ref{GOrela}) and (\ref{o01oij}), we obtain
\begin{equation}
  \Gamma^{0}_{00} = 0 \ , \quad
  \Gamma^{0}_{i0}|_{\gamma_{\tau}} = \Gamma^{i}_{00}|_{\gamma_{\tau}} = a^{i} \ , \quad
  \Gamma^{i}_{j0}|_{\gamma_{\tau}} = - \omega_{k} \epsilon^{0ijk}  \ .  \label{cch}
\end{equation}
We see that the proper reference frame is not a local inertial frame anymore.
Furthermore, considering a spacelike geodesic equation along $\gamma_{s}$,
\begin{equation}
  \frac{d^{2} x^{\mu}}{ds^{2}} 
   + \Gamma^{\mu}_{\alpha\beta} \frac{dx^{\alpha}}{ds} \frac{dx^{\beta}}{ds}  = 0  \ , \label{gams2}
\end{equation}
we can deduce
\begin{equation}
  \Gamma^{\mu}_{ij}(x^{0}=\tau, x^{i}=\alpha^{i}s) \, \alpha^{i} \alpha^{j} = 0 \ .
\end{equation}
Especially, at $s=0$, we obtain 
\begin{equation} 
  \Gamma^{\mu}_{ij}|_{\gamma_{\tau}} = 0 \ .   \label{cch2}
\end{equation}
From Eqs.\,(\ref{cch}), (\ref{cch2}) and the relation between the metric and the Christoffel symbol
\begin{equation}
  g_{\mu\nu,\lambda} =  g_{\mu\alpha} \Gamma^{\alpha}_{\nu\lambda} 
                      + g_{\nu\alpha} \Gamma^{\alpha}_{\mu\lambda} \ , \label{1528}
\end{equation}
the first order derivative of the metric reads 
\begin{eqnarray}
  g_{\mu\nu,0} &=& 0  \ , \nonumber \\
  g_{00,i} &=& -2 a^{i} \ , \nonumber \\
  g_{0i,j} &=& - \omega_{k} \epsilon^{0ijk}    \ ,  \nonumber \\
  g_{ij,k} &=& 0  \ ,
\end{eqnarray}
along the timelike curve $\gamma_{\tau}$.

Next, we evaluate the second order derivatives of the metric.
Differentiating Eqs.\,(\ref{cch}) and (\ref{cch2}) with respect to $\tau$, we get 
\begin{eqnarray}
 && \Gamma^{0}_{00,0}|_{\gamma_{\tau}} = \Gamma^{\mu}_{ij,0}|_{\gamma_{\tau}} = 0 \ ,   \nonumber \\
 && \Gamma^{0}_{i0,0}|_{\gamma_{\tau}} = \Gamma^{i}_{00,0}|_{\gamma_{\tau}} 
        = \dot{a}^{i}  \ ,  \nonumber \\
 && \Gamma^{i}_{j0,0}|_{\gamma_{\tau}} = - \dot{\omega}_{k} \epsilon^{0ijk}  \ ,  \label{chch}
\end{eqnarray}
where a dot represents a derivative with respect to $\tau$.
On the geodesic $\gamma_{\tau}$, from the definition of the Riemann tensor, we find
\begin{equation}
  \Gamma^{\mu}_{\nu0,\lambda}
     = R^{\mu}_{\nu\lambda 0} + \Gamma^{\mu}_{\nu \lambda,0}
       - \Gamma^{\mu}_{\lambda\alpha}\Gamma^{\alpha}_{\nu 0} 
       + \Gamma^{\mu}_{0\alpha}\Gamma^{\alpha}_{\nu \lambda} 
        \ .  \label{CRre}
\end{equation}
Substituting Eqs.\,(\ref{chch}) into Eq.\,(\ref{CRre}), we can deduce
\begin{eqnarray}
  \Gamma^{0}_{00,i}|_{\gamma_{\tau}} &=& \dot{a}^{i} + a^{j}\omega^{k} \epsilon^{0ijk} \ , \nonumber \\
  \Gamma^{0}_{i0,j}|_{\gamma_{\tau}} &=& R^{0}_{ij0}|_{\gamma_{\tau}} - a^{i} a^{j} \ , \nonumber \\
  \Gamma^{i}_{00,j}|_{\gamma_{\tau}} &=& R^{i}_{0j0}|_{\gamma_{\tau}} - \dot{\omega}_{k}\epsilon^{0ijk}
      + a^{i} a^{j} + \omega^{i}\omega^{j} - \delta_{ij} \omega^{k}\omega_{k}  \ , \nonumber \\
  \Gamma^{i}_{j0,k}|_{\gamma_{\tau}} &=& R^{i}_{jk0}|_{\gamma_{\tau}} + a^{j}\omega_{l} \epsilon^{0ikl} \ .
  \label{GGGG}
\end{eqnarray}
In order to obtain an expression for $\Gamma^{\mu}_{ij,k}|_{\gamma_{\tau}}$,
one can utilize a geodesic deviation equation for $\gamma_{s}$ and 
the procedure is completely the same as that in the construction of the Fermi normal coordinates.
Thus, the result is given by Eq.\,(\ref{Gijk}):
\begin{equation}
  \Gamma^{\mu}_{ij,k}|_{\gamma_{\tau}} =
    - \frac{1}{3}\left( R^{\mu}_{ijk}|_{\gamma_{\tau}} + R^{\mu}_{jik}|_{\gamma_{\tau}} \right) \ . 
    \label{Gijk22}
\end{equation}

Finally, we express the second order derivative of the metric by 
the Christoffel symbols and their first derivatives, and then 
relations between the second derivatives of the metric and the Riemann tensor are obtained.
%
%
Differentiating Eq.\,(\ref{1528}) with respect to $x^{\sigma}$, 
we obtain a relation
\begin{equation}
  g_{\mu\nu,\lambda\sigma}|_{\gamma_{\tau}} = 
      \eta_{\mu\alpha} \Gamma^{\alpha}_{\nu\lambda ,\sigma}|_{\gamma_{\tau}}
    + \eta_{\nu\alpha} \Gamma^{\alpha}_{\mu\lambda ,\sigma}|_{\gamma_{\tau}} 
    +  g_{\mu\alpha,\sigma}|_{\gamma_{\tau}}   \Gamma^{\alpha}_{\nu\lambda}|_{\gamma_{\tau}}
    +  g_{\nu\alpha,\sigma}|_{\gamma_{\tau}}   \Gamma^{\alpha}_{\mu\lambda}|_{\gamma_{\tau}} 
    \ .  \label{metmet2}
\end{equation}
Using Eqs.\,(\ref{GGGG}) and (\ref{Gijk22}) in Eq.\,(\ref{metmet2}),
we  can deduce the following equations:
\begin{eqnarray}
  g_{\mu\nu,00} &=& 0  \ , \nonumber \\
  g_{00,0i} &=& -2 \dot{a}^{i} \ , \nonumber \\
  g_{00,ij} &=& -2 R^{0}_{ij0} - 2 a^{i} a^{j} - 2 \omega^{i}\omega^{j} 
                + 2 \delta_{ij} \omega^{k}\omega_{k} \ ,   \nonumber \\
  g_{0i,0j} &=& \dot{\omega}_{k} \epsilon^{0ijk}  \ ,  \nonumber \\
  g_{0i,jk} &=& -\frac{2}{3} 
  \left( R_{0jik}|_{\gamma_{\tau}}  + R_{0kij}|_{\gamma_{\tau}} \right)   \ ,  \nonumber \\
  g_{ij,0k} &=& 0 \ , \nonumber \\
  g_{ij,kl} &=& - \frac{1}{3}
  \left( R_{ikjl}|_{\gamma_{\tau}} + R_{iljk}|_{\gamma_{\tau}} \right)  \ .
\end{eqnarray}
Therefore,
in a proper reference coordinate system, up to the quadratic order of the coordinates, 
the metric is given by
\begin{eqnarray}
  g_{00} &=& - 1 -2 a_{i} x^{i} - \left( a^{i}x^{i} \right)^{2}
             - \left( \omega^{i}x^{i} \right)^{2} + \omega^{i}\omega^{i} x^{j}x^{j}
             - R_{0i0j}|_{\gamma_{\tau}} x^{i} x^{j} \ ,  \label{met011} \\
  g_{0i} &=& - \omega_{k} \epsilon^{0ijk} x^{j}
             -\frac{2}{3} R_{0jik}|_{\gamma_{\tau}} x^{j} x^{k} \ ,  \label{met022} \\
  g_{ij} &=& \delta_{ij} - \frac{1}{3} R_{ikjl}|_{\gamma_{\tau}} x^{k} x^{l} \ .  \label{met033}
\end{eqnarray}

We see that the effects of the Earth enter even at the linear order of
$a^{i}$ and $\omega^{i}$.
However, we can neglect these effects. 
For example, assuming the scale of 
the experimental apparatus to be $x^{i} \sim 1\, {\rm m}$ and 
using the values $a^{i} \sim 9.8 \, {\rm m/s^{2}}$, 
$\omega^{i} \sim 2.0 \times 10^{-7} \, {\rm rad/s}$,
we can estimate
$a^{i}x^{i} \sim 1.1 \times 10^{-16}$ and 
$\omega^{i}x^{i} \sim 6.7 \times 10^{-16}$.
These small corrections are negligible in experiments because
 they are small and their effects are static.
Indeed, the effects of the earth are negligible in magnon experiments
because we utilize a phenomenon of resonance between gravitational waves and 
magnons to detect gravitational waves.
Therefore, we 
use the Fermi normal coordinates approximately for an observer on the earth.
\section{Expansion formula for $e^{iS} H e^{-iS}$ and $\left( \frac{\partial}{\partial t} e^{iS} \right)  e^{-iS}$ } \label{Lie}

Let us introduce a parameter $\lambda$ by
\begin{equation}
  f(\lambda,S) \equiv e^{i\lambda S} H e^{-i\lambda S}  \ .
\end{equation}
We set $\lambda = 1$ after the calculations.
Expanding it with respect to $\lambda$, we obtain
\begin{equation}
  f(\lambda,S) = \sum_{n=0}^{\infty} \frac{\lambda^{n}}{n!} 
                  \left( \frac{\partial^{n} f(\lambda,S)}{\partial \lambda^{n}} \right)_{\lambda=0}  \ .
\end{equation}
We find that 
\begin{eqnarray}
  \frac{\partial f(\lambda,S)}{\partial \lambda} &=& 
      e^{i\lambda S} i \big[ S, H \big]  e^{-i\lambda S} \ , \nonumber \\
  \frac{\partial^{2} f(\lambda,S)}{\partial \lambda^{2}} &=& 
      e^{i\lambda S} i^{2} \big[ S, \big[ S, H \big] \big]  e^{-i\lambda S}  \ ,   \\
      &\vdots&  \nonumber \\
  \frac{\partial^{n} f(\lambda,S)}{\partial \lambda^{n}} &=&
      e^{i\lambda S} i^{n} 
       \big[ S, \big[ S, \ldots  ,\big[ S, H \big] \big] \cdots \big] \big]  e^{-i\lambda S}  \ . \nonumber 
\end{eqnarray}
Therefore, one can deduce
\begin{eqnarray}
  e^{iS} H e^{-iS} &=& f(1,S)  \nonumber \\
                   &=& \sum_{n=0}^{\infty} \frac{i^{n}}{n!} 
                        \big[ S, \big[ S, \ldots  ,\big[ S, H \big] ] \cdots \big] \big] \ . \label{cam}
\end{eqnarray}
The formula (\ref{cam}) is called the Campbell-Baker-Hausdorff formula.

Next, let us consider the expansion of $\left( \frac{\partial}{\partial t} e^{iS} \right)  e^{-iS}$
in powers of $S$.
Again we introduce a parameter $\lambda$ and expand it with respect to $\lambda$ as
\begin{eqnarray}
  g(\lambda,S) &=& \left( \frac{\partial}{\partial t} e^{i\lambda S} \right)  e^{-i\lambda S}  \nonumber \\
               &=& \sum_{n=0}^{\infty} \frac{\lambda^{n}}{n!} 
                  \left( \frac{\partial^{n} g(\lambda,S)}{\partial \lambda^{n}} \right)_{\lambda=0}  \ .
\end{eqnarray}
We see that 
\begin{eqnarray}
  \frac{\partial g(\lambda,S)}{\partial \lambda} &=& 
    e^{i\lambda S}  \,  i \dot{S} \,  e^{-i\lambda S} \ , \nonumber \\
  \frac{\partial^{2} g(\lambda,S)}{\partial \lambda^{2}} &=& 
       e^{i\lambda S} i^{2} \big[ S, \dot{S} \big]  e^{-i\lambda S}  \ ,   \\  \label{A6}
      &\vdots&  \nonumber \\
  \frac{\partial^{n+1} f(\lambda,S)}{\partial \lambda^{n+1}} &=&
      e^{i\lambda S} i^{n+1} 
       \big[ S, \big[ S, \ldots  ,\big[ S, \dot{S} \big] \big] \cdots \big] \big]  e^{-i\lambda S}  \ . \nonumber 
\end{eqnarray}
Note that the right-hand side of the last equation has $n$ pieces of $S$.
Hence, we finally arrive at
\begin{eqnarray}
  \left( \frac{\partial}{\partial t} e^{iS} \right)  e^{-iS}
     &=& g(1,S)  \nonumber \\
     &=& \sum_{n=0}^{\infty} \frac{i^{n+1}}{(n+1)!} 
                   \big[ S, \big[ S, \ldots  ,\big[ S, \dot{S} \big] \big] \cdots \big] \big] \ . \label{expanS}
\end{eqnarray}
\bibliography{non_rela}

\end{document}